\documentclass[12pt]{article}
\input xy
\xyoption{all}

\usepackage{amsmath,amssymb,amsfonts}
\usepackage{color}
\usepackage{hyperref}

\begin{document}

\vspace*{0.3in}

\begin{center}

{\large\bf D-brane probes, branched double covers, }

{\large\bf and noncommutative resolutions}

\vspace{0.5in}

Nicolas M. Addington$^{1}$, Edward P. Segal$^2$, Eric R. Sharpe$^3$

\begin{tabular}{cc}
{ \begin{tabular}{c}
$^1$ Mathematics Department\\
Duke University, Box 90320\\
Durham, NC  27708-0320
\end{tabular} }
&
{ \begin{tabular}{c}
$^3$ Physics Department\\
Robeson Hall, 0435\\
Virginia Tech \\
Blacksburg, VA  24061
\end{tabular} } \\
{ \begin{tabular}{c}
$^2$ Mathematics Department\\
Imperial College\\
London SW7 2AZ
\end{tabular} }
& { $\,$ }
\end{tabular}

{\tt adding@math.duke.edu}, \\
{\tt edward.segal04@imperial.ac.uk}, \\
{\tt ersharpe@vt.edu}\\

\end{center}

This paper describes D-brane probes of theories arising in abelian
gauged linear sigma models (GLSMs) describing branched double covers
and noncommutative resolutions thereof, via nonperturbative effects rather than
as the critical locus of a superpotential.  As these theories can be
described as IR limits of Landau-Ginzburg models, technically this paper
is an exercise in utilizing (sheafy) matrix factorizations.
For Landau-Ginzburg models which are believed to flow in the IR to
smooth branched double covers, our D-brane probes recover the 
structure of the branched double cover (and flat nontrivial $B$ fields),
verifying previous results.
In addition to smooth branched double covers, the same class of
Landau-Ginzburg models is also believed to sometimes flow to
`noncommutative resolutions' of singular spaces.  
These noncommutative resolutions are abstract conformal field theories
without a global geometric description, but D-brane probes perceive them as
a non-K\"ahler small resolution of a singular Calabi-Yau.  We conjecture that
such non-K\"ahler resolutions are typical in D-brane probes of such
theories.

\begin{flushleft}
November 2012
\end{flushleft}

\newpage

\tableofcontents

\newpage

\section{Introduction}

For many years it was thought that gauged linear sigma models (GLSMs)
could only
describe geometries built as global 
complete intersections, and that all geometric
phases of GLSMs were birational to one another.  This lore has been
contradicted by a series of recent papers 
\cite{hhpsa,horitong,ron-sharpe,cdhps,horispr11,jklmr} describing
examples of GLSMs in which
\begin{itemize}
\item geometries are built via nonperturbative effects 
\cite{hhpsa,horitong,ron-sharpe,cdhps}, instead of
perturbatively as the critical locus of a superpotential,
\item geometries not described as global complete intersections are
constructed (sometimes via nonperturbative effects, and more recently,
sometimes perturbatively \cite{horispr11,jklmr}),
\item geometric phases are not birational to one another.
(It is now believed that phases are instead related by `homological
projective duality' \cite{ron-sharpe,cdhps,kuz1,kuz2,kuz3,bfk}.)
\end{itemize}

In addition, and perhaps even more importantly,
new (2,2) SCFT's describing examples of `noncommutative
resolutions' were also constructed \cite{cdhps}, as IR limits of certain GLSMs.
Since the term `noncommutative geometry' has been applied to describe a variety
of situations in string theory, let us take a moment to clarify.
One popular usage is as a way of
understanding D-branes in a certain coupling limit, as in
\cite{seib-wit}; another is in \cite{rw}; but neither of these usages
is precisely what we have in mind.  Instead, the noncommutative
resolutions described in \cite{cdhps} are closed string SCFT's with
open string sectors (described via matrix factorizations
in UV Landau-Ginzburg models) realizing structures described
in 
{\it e.g.} \cite{kuz1,kuz2,kuz3,bfk,vdb1,vdb2,vdb3,blb,dao,bl}.
Briefly, the closed string theories were understood as abstract
conformal field theories, which had
a geometric interpretation in some local patches, but not globally.
In other patches, the open string theory could only be understood
as a noncommutative or nc resolution of a naive singularity (not present
physically in the abstract conformal field theory), described mathematically
in the references above.

This paper will study D-brane probes of the various geometries,
both ordinary and noncommutative, arising in \cite{cdhps}.
That paper described nonlinear sigma models on smooth branched
double covers and nc resolutions of singular branched double covers,
arising via nonperturbative effects, and so D-brane probes
provide both a useful check
of the results as well as insight into the properties of the new CFT's. 

For GLSM's describing nonlinear sigma models on smooth branched double
covers,
D-brane probes will recover the same results as \cite{cdhps} -- we will
get an alternative derivation of the results described in that paper.

For noncommutative resolutions, our methods will provide some novel
physical insights.  Specifically, in our analysis, D-brane probes
of noncommutative resolutions will see a 
non-K\"ahler small resolution
of the singularities.  As the D-brane
probe moduli space is not\footnote{
It is not believed to possess a generalized Calabi-Yau structure in the
sense of Hitchin's generalized complex geometry.
} compatible with (2,2) worldsheet supersymmetry,
we do not interpret it as the
closed string sector.  (Indeed, D-brane probes often see different
geometries from closed string sectors; for another example, see
\cite{dgm,dks} for a discussion of how D-brane probes of orbifolds see
resolutions of quotient spaces, instead of the closed string's
target-space geometry.)
We will very briefly discuss how
this may generalize to other cases.

In all cases, we can understand the results of \cite{cdhps} in terms
of low-energy behaviors of certain Landau-Ginzburg theories, so
at a technical level this paper will concern D-brane probes of certain
Landau-Ginzburg models.  D-branes in Landau-Ginzburg theories
are described by `matrix factorizations,' and our D-brane probe arguments
will utilize `sheafy matrix factorizations,' 
{\it i.e.},
matrix factorizations supported over subvarieties of the space, rather than
complexes of locally-free sheaves.

Other examples of noncommutative resolutions pertinent to 
physics in different ways have appeared
in \cite{vb,sz,cv,pd}.  We will focus on D-brane probes of the nc spaces
appearing in \cite{cdhps} in this paper.

We begin in section~\ref{sect:glsm:nc} with an overview of the methods and
results of \cite{cdhps}.  We describe the particular GLSMs that flow in the
IR to branched double covers, both smooth and nc resolutions.
In this paper we construct D-brane probes, which we will realize
via sheafy matrix factorizations in intermediate-energy Landau-Ginzburg
theories.  Ordinary matrix factorizations are now a staple of the
literature, but sheafy matrix factorizations are still relatively new,
so in section~\ref{sect:matrix-fact-sub} we review some pertinent properties
of sheafy matrix factorizations, beginning with their physical derivation
and running through both a mathematical description of their RG flow,
as well as an analysis of what it means for a sheafy matrix factorization
to be `point-like' (which plays a crucial role in D-brane probes).

In section~\ref{sect:smooth} we apply that technology to compute
D-brane probe moduli spaces of GLSMs and Landau-Ginzburg models that
flow in the IR to nonlinear sigma models on smooth branched double covers.
That double cover structure is realized via nonperturbative effects,
and so D-brane probes can provide a useful consistency check of
other analyses.  We begin the section with a brief overview of how
D-brane probes work in Landau-Ginzburg models that flow to smooth
manifolds realized perturbatively, instead of nonperturbatively, and then
go on to describe various examples.  Most of our analysis is local,
but we conclude the section by describing global gluing issues that exist,
and how they sometimes
predict the presence of a topologically nontrivial $B$ field.

In section~\ref{sect:nc} we construct moduli spaces of D-brane probes
of Landau-Ginzburg theories that flow in the IR to nc resolutions of
singular branched double covers.
As outlined above, in examples, 
we find that the D-brane probe moduli spaces are
(non-K\"ahler) small resolutions of the singular branched double covers. 
We also outline more general statements which suggest that D-brane probes
of nc resolutions will always see (typically non-K\"ahler) small
resolutions.

We also provide a few appendices to make this paper self-contained.
Appendix~\ref{sect:tech} concerns mathematical technology for
manipulating sheafy matrix factorizations and computing Ext groups
between them.  Appendix~\ref{app:pointlike} gives a general argument
for why the prototype for `point-like' matrix factorizations is, indeed,
pointlike.

\section{Branched double covers, nc resolutions, and GLSM's}  
\label{sect:glsm:nc}

In this section, we review the results of \cite{cdhps}, in which
branched double covers and
noncommutative resolutions were realized in abelian GLSMs.

The simplest example discussed in \cite{cdhps} was the
GLSM for the complete intersection Calabi-Yau ${\mathbb P}^3[2,2]$.
The superpotential for this theory is of the form
\begin{displaymath}
W \: = \: \sum_a p_a G_a(\phi) \: = \:
\sum_{ij} \phi_i \phi_j A^{ij}(p)
\end{displaymath}
where the $\phi$'s act as homogeneous coordinates on ${\mathbb P}^3$,
the $G_a$'s are the two quadrics, and $A^{ij}$ is a symmetric $4 \times 4$
matrix with entries linear in the $p$'s, determined by the $G_a$'s.

At the Landau-Ginzburg point of this theory, where the $p_a$ are not
all zero, the superpotential acts as a mass matrix for $\phi$'s.
Naively, this is problematic:  we are left with a theory containing
only $p$'s, which looks like a sigma model on ${\mathbb P}^1$, which cannot
possibly be Calabi-Yau.  However, a closer analysis reveals subtleties.
First, since the $p$'s are charge 2, there is a trivially-acting
${\mathbb Z}_2$ here (technically, a ${\mathbb Z}_2$ gerbe structure),
which physics interprets \cite{hhpsa} as a double cover.
Second, the mass matrix $A^{ij}(p)$ has zero eigenvalues along the 
degree four hypersurface $\{ \det A = 0\}$.  
With a bit of further analysis discussed in
\cite{cdhps}, one argues that this flows in the IR to a nonlinear sigma
model on 
a branched
double cover of ${\mathbb P}^1$, branched over a degree four
hypersurface --
an example of a Calabi-Yau.  In fact, both ${\mathbb P}^3[2,2]$
and the branched double cover are elliptic curves.

Analogous analyses apply to many other examples.  The next simplest
involves the GLSM for ${\mathbb P}^5[2,2,2]$, which is a K3 surface.
Its Landau-Ginzburg point is interpreted as a branched double cover
of ${\mathbb P}^2$, branched over a degree six locus, which is another
K3.

If the projective space is even-dimensional, then the analysis is
somewhat more complicated.  Examples of this form were outlined in
\cite{cdhps}, but not studied in as much detail.  Briefly, in such
cases, instead of a branched double cover (which turns out not to exist
globally), instead one gets a single cover with a locus of
${\mathbb Z}_2$ orbifolds replacing what would have been the branch locus,
$\{ \det A = 0 \}$.

For some examples of this form, this is the complete story, and the 
Landau-Ginzburg model is indeed believed to RG flow to a 
branched double cover.  However, in many higher dimensional examples,
there are further wrinkles to the story.  Consider, for example,
the GLSM for ${\mathbb P}^7[2,2,2,2]$.  This is a Calabi-Yau, and the
analysis above suggests that its Landau-Ginzburg point should 
flow in the IR to a nonlinear sigma model on
a branched double cover of ${\mathbb P}^3$,
branched over a degree eight locus.
However, there is a problem with this interpretation: 
the branched double cover above is singular, but the GLSM
is smooth.  This was interpreted in \cite{cdhps} as a `noncommutative
resolution' of the singularities, one specifically discussed in 
\cite{kuz2}, based on the fact that 
the corresponding Landau-Ginzburg model, intermediate along RG flow,
had matrix factorizations in precise correspondence with sheaves defining
the noncommutative resolution in \cite{kuz2}.  
(We will describe the details of that
noncommutative resolution shortly.)

In passing, the geometries above (the complete intersection
${\mathbb P}^7[2,2,2,2]$, the branched double cover of ${\mathbb P}^3$ branched
along the octic) are not birational to one another, violating unproven old
lore concerning geometric limits of GLSM's.  Instead, it was proposed
in \cite{cdhps} that they are related by ``homological projective duality''
\cite{kuz1,kuz2,kuz3} which predicts relationships of exactly the form above,
including noncommutative resolutions.

Let us review the highlights so far:
\begin{itemize}
\item In this GLSM, geometry is realized in a rather novel fashion,
via nonperturbative effects.
\item The resulting (nearly) geometric phases are not birational to one
another.
\item We have a physical realization of a closed string theory
corresponding to a noncommutative resolution of a space -- in other
words, a physical realization of a new kind of (2,2) SCFT in
two dimensions.
\item There exists a mathematical understanding of how to relate
both spaces and noncommutative resolutions appearing at different
ends of GLSM K\"ahler moduli spaces, known as ``homological projective
duality.''
\end{itemize}
The first two points were also described in nonabelian GLSM's in
\cite{horitong,ron-sharpe,horispr11}.  Their different geometric phases
are also related by homological projective duality.
(In addition, we have been told \cite{tonypriv} that even in the
examples of \cite{horitong,horispr11}, there are special points in the
moduli space at which the dual theories are mathematically singular,
but should, according to the predictions of homological projective duality,
be understood as noncommutative resolutions.)
 
Now, let us describe in detail the noncommutative resolution of a 
branched double cover appearing at one end of the GLSM K\"ahler moduli
space for ${\mathbb P}^7[2,2,2,2]$.
Mathematically \cite{kuz2}, it is described by a noncommutative variety
$({\mathbb P}^3,{\cal B})$, where ${\cal B} \in {\rm Coh}({\mathbb P}^3)$
is the sheaf of even parts of Clifford algebras associated with the
universal quadric defined by the matrix $A^{ij}(p)$.  (In other words,
each point on ${\mathbb P}^3$ defines a symmetric $8 \times 8$ matrix
$A^{ij}(p)$ (up to overall rescaling), to which we can associate a 
Clifford algebra.)
Equivalently, we could consider the double cover $f: Z \rightarrow 
{\mathbb P}^3$,
together with a sheaf of algebras ${\cal A} \rightarrow Z$ for which
$f_* {\cal A} = {\cal B}$.

Intuitively, we can think about how this nc resolution resolves a
singularity as follows.
The nc resolution adds more sheaves
over the singularity, as if the singularity were resolved. 
For example, if one small resolves a conifold singularity,
the resolution has extra sheaves supported over the inserted ${\mathbb P}^1$
than the original singularity has supported at the singularity.
Part of the story here is that a nc resolution adds extra sheaves
to similarly `resolve' the singularity (though the result lacks a
simple geometric understanding).
 
Physically, we can understand how the GLSM realizes this structure as follows.
(This analysis was also presented in \cite[section 2.6.2]{cdhps}.)
At the Landau-Ginzburg point, we can go to an intermediate point in RG flow,
where the theory is a Landau-Ginzburg model on
\begin{displaymath}
{\rm Tot}\left( {\cal O}(-1)^8 \: \stackrel{\pi}{\longrightarrow}
\: {\mathbb P}^3_{[2,2,2,2]} \right)
\end{displaymath}
with superpotential
\begin{displaymath}
W \: = \: \sum_{ij} \phi_i \phi_j A^{ij}(p)
\end{displaymath}
(see \cite[section 4]{alg22} for more examples of this sort of analysis),
or, more simply but slightly less accurately,
a ${\mathbb Z}_2$ orbifold of a Landau-Ginzburg model
on ${\mathbb P}^3$ with a quadric superpotential.

Now, matrix factorizations for quadratic superpotentials were
studied in \cite{kapli}, where it was argued that D0-branes in such
Landau-Ginzburg models have a Clifford algebra structure, and other branes
are naturally acted upon by the D0-branes.  In the present case, applying
a Born-Oppenheimer analysis, we have immediately that all D-branes should
be modules over a sheaf of Clifford algebras, and taking into account the
${\mathbb Z}_2$ orbifold structure, modules over a sheaf of {\it even} parts
of Clifford algebras, which is precisely the sheaf ${\cal B}$ appearing
in the mathematical definition of the noncommutative resolution. 
(A mathematically rigorous discussion of matrix factorizations
in this example has also been worked out
\cite{kuzpriv}.)

Broadly speaking, this sort of structure should be fairly universal in
GLSM's describing complete intersections of quadric hypersurfaces, and,
indeed, a number of other examples were discussed in \cite{cdhps}.

In this paper we shall use D-brane probes to study both the branched
double covers and these
noncommutative resolutions.  For Landau-Ginzburg models believed
to flow to smooth branched double covers, our D-brane probe analysis will
confirm this interpretation.  For Landau-Ginzburg models believed to
flow to noncommutative resolutions, our D-brane probe analysis will give
insight into the corresponding closed string conformal field theories.

Our analysis of D-brane probes revolves around nontraditional
matrix factorizations, involving sheaves with support on positive
codimension subvarieties, so in the next section we will review such
`sheafy' matrix factorizations.

\section{Sheafy matrix factorizations}
\label{sect:matrix-fact-sub}

Our D-brane probes of Landau-Ginzburg models will involve
`sheafy' matrix factorizations, involving pairs of maps between
coherent sheaves supported over subvarieties of the base.

Most matrix factorizations discussed in the literature involve
pairs of maps between bundles, not sheaves. However, as we shall review,  
the Warner problem only exists for Neumann boundary conditions, 
so for example, matrix factorizations supported over a 
submanifold are allowed. The sheafy matrix factorizations that we will utilize
have been discusssed in a few recent
mathematics
papers (see {\it e.g.} \cite{bfk,pos1,orlov11,isik10,lin-pomerleano,bw}), 
but not often, and we will need a number of properties of
these matrix factorizations.  Since they are still not entirely common,
and we will need a number of results, in this section we will 
review their properties.  We begin with a physics analysis, rederiving
matrix factorizations in this more general context (see also
one of the authors' lecture notes \cite{snowbird}), 
and then turn to
mathematical properties of sheafy matrix factorizations.  We use
mathematics as a guide to understand renormalization group flow of
sheafy matrix factorizations in Landau-Ginzburg models, and in doing so
we will uncover some surprising behaviors.

For an orthogonal overview of traditional
matrix factorizations, see \cite{calin1}.
For a more traditional approach to matrix factorizations
and D-brane probes, see for example \cite{paul1,paul2}.

\subsection{Physics analysis}\label{sect:physicsanalysis}

The most general Landau-Ginzburg model (over a space) that one can write
down has the following (bulk) action:
\begin{eqnarray*}
\lefteqn{ 
\frac{1}{\alpha'} \int_{\Sigma} d^2z \Bigg(
%g_{i \overline{\jmath}} \overline{\partial} \phi^i \partial \phi^{\overline{\jm
%ath}} \: + \:
\frac{1}{2} g_{\mu \nu} \partial \phi^{\mu} \overline{\partial} \phi^{\nu}
\: + \: \frac{i}{2} B_{\mu \nu} \partial \phi^{\mu} \overline{\partial}
\phi^{\nu} \: + \:
%i g_{i \overline{\jmath}} \psi_-^{\overline{\jmath}} D_z \psi_-^i \: + \:
%i g_{i \overline{\jmath}} \psi_+^{\overline{\jmath}} D_{\overline{z}} \psi_+^i 
\frac{i}{2} g_{\mu \nu} \psi_-^{\mu} D_z \psi_-^{\nu} \: + \:
\frac{i}{2} g_{\mu \nu} \psi_+^{\mu} D_{\overline{z}} \psi_+^{\nu}
} \\
& & \hspace*{1.0in}\: + \:
R_{i \overline{\jmath} k \overline{l}} \psi_+^i \psi_+^{\overline{\jmath}} 
\psi_-^k
\psi_-^{\overline{l}}
%\right. 
%\\
%& & \hspace*{2.5in}  %\left. 
\: 
- \: g^{i \overline{\jmath}} \partial_i W \partial_{\overline{\jmath}}
\overline{W} \: + \: i \psi_+^i \psi_-^j D_i \partial_j W \: + \:
i \psi_+^{\overline{\imath}} \psi_-^{\overline{\jmath}} D_{\overline{\imath}}
\partial_{\overline{\jmath}} \overline{W} \Bigg)
\end{eqnarray*}
where $W$ is the superpotential,
a holomorphic function over the target space $X$,
and
\begin{displaymath}
D_i \partial_j W \: = \: \partial_i \partial_j W \: - \:
\Gamma_{ij}^k \partial_k W
\end{displaymath}
%The bosonic potential is of the form $\sum_i | \partial_i W |^2$.
The action possesses the supersymmetry transformations:
\begin{eqnarray*}
\delta \phi^i & = & i \alpha_- \psi_+^i \: + \: i \alpha_+ \psi_-^i \\
\delta \phi^{\overline{\imath}} & = & i \tilde{\alpha}_- \psi_+^{
\overline{\imath}} \: + \: i \tilde{\alpha}_+ \psi_-^{\overline{\imath}} \\
\delta \psi_+^i & = & - \tilde{\alpha}_- \partial \phi^i \: - \:
i \alpha_+ \psi_-^j \Gamma^i_{j m} \psi_+^m \: + \:
\alpha_+ g^{i \overline{\jmath}} \partial_{\overline{\jmath}} \overline{W}\\
\delta \psi_+^{\overline{\imath}} & = & - \alpha_- \partial 
\phi^{\overline{\imath}}
\: - \: i \tilde{\alpha}_+ \psi_-^{\overline{\jmath}} 
\Gamma^{\overline{\imath}}_{\overline{\jmath} \overline{m} }
\psi_+^{\overline{m}} 
\: + \: \tilde{\alpha}_+ g^{\overline{\imath} j} \partial_j W \\
\delta \psi_-^i & = & - \tilde{\alpha}_+ \overline{\partial} \phi^i \: - \:
i \alpha_- \psi_+^j \Gamma^i_{j m} \psi_-^m 
\: - \: \alpha_- g^{i \overline{\jmath}} \partial_{\overline{\jmath}}
\overline{W} \\
\delta \psi_-^{\overline{\imath}} & = & - \alpha_+ \overline{\partial}
\phi^{\overline{\imath}} \: - \: 
i \tilde{\alpha}_- \psi_+^{\overline{\jmath}} 
\Gamma^{\overline{\imath}}_{\overline{\jmath} \overline{m}}
\psi_-^{\overline{m}} 
\: - \: \tilde{\alpha}_- g^{\overline{\imath} j} \partial_j W
\end{eqnarray*}
(See \cite{alg22} for a discussion of closed string
A- and B-twisted Landau-Ginzburg
models on nontrivial spaces.  In this section we will focus on open strings.)

Under a supersymmetry transformation, the bulk action picks up the following
total derivative terms:
\begin{eqnarray}   \label{warnerterms}
\lefteqn{
\frac{1}{\alpha'} \int_{\Sigma} d^2z \Bigg[
\partial \left( - \frac{i}{2} \alpha_- \partial_{\overline{\jmath}}
\overline{W} \psi_-^{\overline{\jmath}} \right) \: + \:
\overline{\partial} \left( \frac{i}{2} \alpha_+ \partial_{\overline{\jmath}}
\overline{W} \psi_+^{\overline{\jmath}} \right)
}
\\
& & \hspace*{1.0in} \: + \:                       
\partial \left( -\frac{i}{2} \tilde{\alpha}_- \partial_i W \psi_-^i \right)
\: + \:           
\overline{\partial} \left( \frac{i}{2} \tilde{\alpha}_+ \partial_i W \psi_+^i \right)             
\Bigg] \nonumber
\end{eqnarray}
If we take $\Sigma$ to be the upper half-plane for simplicity, so that
\begin{displaymath}
\int_{\Sigma} d^2z \, \partial \: = \: \frac{1}{2i} 
\int_{\partial \Sigma} dx, \: \: \:
\int_{\Sigma} d^2z \, \overline{\partial} \: = \:
- \frac{1}{2i} \int_{\partial \Sigma} dx
\end{displaymath}
then we see the total derivative terms above become
\begin{eqnarray*}
\frac{1}{\alpha'} \frac{1}{2i} \int_{\partial \Sigma} dx \left[
- \frac{i}{2} \alpha \partial_{\overline{\imath}} \overline{W}
\psi_-^{\overline{\imath}}
\: - \: \frac{i}{2} \alpha \partial_{\overline{\imath}}
\overline{W}
\psi_+^{\overline{\imath}} \right]
& = & - \frac{1}{\alpha'} \frac{1}{4} \int_{\partial \Sigma} dx \left[
\alpha \partial_{\overline{\imath}} \overline{W} \left(
\psi_+^{\overline{\imath}} \: + \: \psi_-^{\overline{\imath}} \right)
\right] \\ 
\frac{1}{\alpha'} \frac{1}{2i} \int_{\partial \Sigma} dx \left[
- \frac{i}{2} \tilde{\alpha} \partial_i W \psi_-^i \: - \:
\frac{i}{2} \tilde{\alpha} \partial_i W \psi_+^i \right]
& = &
- \frac{1}{\alpha'} \frac{1}{4} \int_{\partial \Sigma} dx \left[
\tilde{\alpha} \partial_i W \left( \psi_+^i \: + \: \psi_-^i \right) \right]
\end{eqnarray*}
where we have defined $\alpha = \alpha_- = \alpha_+$,
$\tilde{\alpha} = \tilde{\alpha}_+ = 
\tilde{\alpha}_-$, using an identity that exists for both
Dirichlet and Neumann boundary conditions\footnote{
Our analysis neglects an aspect of Chan-Paton fields coupling to a
bundle with nonzero curvature, namely that such curvature modifies
boundary conditions \cite{acny}, which often plays an important 
role, such as in \cite{ks}.  That said, the complete role of such curvatures
is only understood when the Chan-Paton factors couple to line bundles;
the resulting modifications induced by nonabelian gauge fields do not seem
to be currently understood, essentially because of technical issues with
path-ordered exponentials in this context.  
For simplicity, in this section, we shall
implicitly specialize to the case that curvatures are trivial,
and we will assume later 
that the results obtained in this special case generalize
in the obvious fashion, following the same pattern as most other physics
papers on derived categories.
}.

In the special case of Dirichlet boundary conditions,
$\psi_+^{\mu} = - \psi_-^{\mu}$, so we see the terms above cancel out.
However, for Neumann boundary conditions,
$\psi_+^{\mu} = + \psi_-^{\mu}$, and so the terms above do not cancel out.

To solve this problem along Neumann directions, we introduce a boundary
action describing
a brane, antibrane, and tachyons.
The boundary action is then (\cite[section 5.1.2]{hori-lin}, 
\cite[section 4]{kl}, \cite[section 2]{ttu}, \cite{hhp})
\begin{eqnarray*}
\lefteqn{
- \frac{1}{4\alpha'} \int_{\partial \Sigma} dx \left[
h_{\alpha \overline{\alpha}} h_{a \overline{a}} 
\overline{\eta}^{\overline{\alpha} \overline{a}} d \eta^{\alpha a}
\: + \:
i \psi^i \left( \partial_i P_{\alpha a} \right) \eta^{\alpha a}
\: + \: i \psi^{\overline{\imath}} \left(
\partial_{\overline{\imath}} \overline{P}_{\overline{\alpha}
\overline{a}} \right) \overline{\eta}^{\overline{\alpha} \overline{a}} 
\right. } \\
& \hspace*{1in} & + \: 
i \psi^i \left( \partial_i Q^{\alpha a} \right) \overline{\eta}^{
\overline{\alpha} \overline{a}} h_{\alpha \overline{\alpha}}
h_{a \overline{a}} \: + \:
i \psi^{\overline{\imath}} \left( \partial_{\overline{\imath}}
\overline{Q}^{\overline{\alpha} \overline{a}} \right) \eta^{\alpha a}
h_{\alpha \overline{\alpha}} h_{a \overline{a}} \\
& \hspace*{1in} & \hspace*{0.5in}
\left. - i P_{\alpha a} \overline{P}_{\overline{\alpha} \overline{a}}
h^{\alpha \overline{\alpha}} h^{a \overline{a}} \: - \:
i Q^{\alpha a} \overline{Q}^{\overline{\alpha} \overline{a}}
h_{\alpha \overline{\alpha}} h_{a \overline{a}} \right]
\end{eqnarray*}
where $\psi^i = \psi_+^i + \psi_-^i$,
$\psi^{\overline{\imath}} = \psi_+^{\overline{\imath}}
+ \psi_-^{\overline{\imath}}$, and $\eta$, $\overline{\eta}$ are
fermions that only live along the boundary $\partial \Sigma$.
If we let ${\cal E}_0$, ${\cal E}_1$ denote the two
holomorphic vector
bundles 
appearing along
the boundary, then $h_{\alpha \overline{\alpha}}$, $h_{a \overline{a}}$,
respectively, are their hermitian fiber metrics (which we have assumed
constant in stating that the connections vanish).
The boundary fermions $\eta$, $\overline{\eta}$ couple to
${\cal E}_0^{\vee}\otimes {\cal E}_1$ and 
${\cal E}_0 \otimes {\cal E}_1^{\vee}$,
respectively (which is slightly obscured by our notation).
The fields $P_{\alpha a}$, $Q^{\alpha a}$ are holomorphic sections
of ${\cal E}_0^{\vee} \otimes {\cal E}_1$ and 
${\cal E}_0 \otimes {\cal E}_1^{\vee}$.
$P$ and $Q$ are the two tachyons mentioned earlier, connecting the
brane to the anti-brane.

Take the supersymmetry variations of $\phi$, $\psi$ along the boundary
to be the restriction to the boundary of the bulk supersymmetry
transformations, and take the boundary fermions $\eta$, $\overline{\eta}$
to have supersymmetry variations
\begin{eqnarray*}
\delta \eta^{\alpha a} & = &
-i h^{\alpha \overline{\alpha}} h^{a \overline{a}} 
\overline{P}_{\overline{\alpha} \overline{a}} \alpha
\: - \: i Q^{\alpha a} \tilde{\alpha} \\
\delta \overline{\eta}^{\overline{\alpha} \overline{a}} & = &
-i h^{\alpha \overline{\alpha}} h^{a \overline{a}}
P_{\alpha a} \tilde{\alpha} \: - \: i 
\overline{Q}^{\overline{\alpha} \overline{a}} \alpha
\end{eqnarray*}
then the supersymmetry variation of the boundary action above is given by
\begin{equation}  \label{bdvar}
- \frac{1}{4 \alpha'} \int_{\partial \Sigma} dx \left[
- \alpha \psi^{\overline{\imath}} \partial_{\overline{\imath}}
\left( \overline{P}_{\overline{\alpha} \overline{a}}
\overline{Q}^{\overline{\alpha} \overline{a}} \right)
\: - \: 
\tilde{\alpha} \psi^i \partial_i \left(
P_{\alpha a} Q^{\alpha a} \right)  \right]
\end{equation}

Comparing to equation~(\ref{warnerterms}), it is easy to see that
the Warner problem will be solved, the total boundary term in the
supersymmetry variations will vanish, if we choose $P$, $Q$ such that
\begin{displaymath}
P_{\alpha a} Q^{\alpha a} \: = \: W
\end{displaymath}
(up to an overall constant shift).

This is the solution to the Warner problem:  to introduce
two bundles ${\cal E}_0$, ${\cal E}_1$, living on the submanifold defined
by the Dirichlet boundary conditions, together with maps
$P: {\cal E}_0 \rightarrow {\cal E}_1$ and $Q: {\cal E}_1 \rightarrow 
{\cal E}_0$ such that $P \circ Q = W \, \mbox{Id}$, up to a constant shift.
We shall typically denote this data as
\begin{displaymath}
\xymatrix{
{\cal E}_0 \ar@/_/[d]_P \\
{\cal E}_1 \ar@/_/[u]_Q
}
\end{displaymath}
Since $P$ and $Q$ are matrix-valued functions that `factorize' $W$, 
this data is known
as a matrix factorization.
Each such matrix factorization defines a D-brane (or collection thereof)
in the Landau-Ginzburg
model.

For example, skyscraper sheaves over any point define easy examples of
sheafy matrix
factorizations.  If $x$ is any point, then
\begin{displaymath}
\xymatrix{
{\cal O}_x \ar@/_/[d] \\
0 \ar@/_/[u]
}
\end{displaymath}
is an example of a matrix factorization, for the trivial reason
that since $W$ is constant on the support, it suffices to find maps $P$, $Q$
whose composition vanishes.\footnote{There is an issue here - if $x$ and $y$ are distinct points on which $W$ takes different values, then the space of morphisms between the two skyscraper sheaves (see appendix~\ref{sect:app:duals-homs}) will fail to be a complex.  This can be remedied by the
introduction of curved dg structures, but we will instead assume that we have a $\mathbb{C}^\times_R$ symmetry, as we discuss shortly. Then a point $x$ defines a matrix factorization only if it is fixed by this symmetry, which implies that $W(x)=0$.}

In fact it is not necessary to assume that ${\cal E}_0$ and ${\cal E}_1$ are supported on the same submanifold, as long as $W$ is constant along the locus where only one of the two bundles is supported. We can also allow ${\cal E}_0$ and ${\cal E}_1$ to be general coherent sheaves, with different ranks at different points, by stacking D-branes of different dimensions on top on one another.

Next, let us review the possible R symmetries of this theory.
First, for any superpotential over $X$, 
there is a ${\mathbb Z}_2$ R symmetry
under which $\psi \mapsto - \psi$ and $\eta \mapsto - \eta$, and $X$
is invariant.  We will
denote this symmetry as ${\mathbb Z}_2^R$.  This R symmetry distinguishes
branes from antibranes -- one is an invariant eigensheaf under
${\mathbb Z}_2^R$, the other is an anti-invariant eigensheaf.
We shall assume ${\cal E}_0$ is invariant and ${\cal E}_1$ is
anti-invariant.

Sometimes, for some spaces and some superpotentials, the ${\mathbb Z}_2^R$
can be extended to a ${\mathbb C}^{\times}$ symmetry.
We will let ${\mathbb C}^{\times}_R$ denote a 
${\mathbb C}^{\times}$
action on $X$, together with ${\mathbb C}^{\times}$-equivariant structures
on ${\cal E}_0$, ${\cal E}_1$, such that
\begin{enumerate}
\item
the superpotential $W$ has weight\footnote{
Physics only depends upon $dW$, not $W$, so we are free to add a constant
to $W$ without changing the physics.  Unfortunately, the definition above
does not respect this symmetry, as adding a constant would spoil the
quasi-homogeneity of $W$ and hence the first part of the definition.
One way to proceed
would be to define the R symmetry in terms of $dW$ instead of $W$;
instead, in this paper we define $W$ to include whatever constant shift is 
needed
in order for an R symmetry to exist.
} two, 
\item $-1 \in {\mathbb C}^{\times}$ acts trivially on $X$, 
\item
the same $-1 \in {\mathbb C}^{\times}$ generates the universal
${\mathbb Z}_2^R$,
\item in matrix factorizations, $P$ and $Q$ each have weight one.
\end{enumerate}

This definition of a $\mathbb{C}^\times_R$ symmetry was introduced, 
at least into the mathematics literature, in \cite{segal}. 
In local coordinates, we can describe this vector ${\mathbb C}^{\times}_R$
action on the closed string sector as follows:
\begin{eqnarray*}
\delta \phi^i & = & \epsilon X^i \\
\delta \phi^{\overline{\imath}} & = & \epsilon X^{\overline{\imath}} \\
\delta \psi_{\pm}^i & = & - \alpha \epsilon \psi_{\pm}^i \: + \:
\epsilon \psi_{\pm}^j \partial_j X^i \\
\delta \psi_{\pm}^{\overline{\imath}} & = & + \alpha \epsilon \psi_{\pm}^{
\overline{\imath}} \: + \: \epsilon \psi_{\pm}^{\overline{\jmath}}
\partial_{\overline{\jmath}} X^{\overline{\imath}} \\
\delta W & = & 2 \alpha \epsilon W
\end{eqnarray*}
where $X^i$ are the components of a holomorphic Killing vector,
$\epsilon$ a small parameter, and $\alpha$ an arbitrary constant, whose
presence reflects the fact that, if one omits the superpotential,
the ${\mathbb C}^{\times}$ actions on the base and on the fermions
are decoupled, which is the reason that the third axiom above 
suffices to ensure
that the entire ${\mathbb C}^{\times}$, not just a ${\mathbb Z}_2$
subgroup, acts as an R symmetry.

Let us briefly describe an example in detail.
Consider the special case that $X = {\mathbb C}^2$, with coordinates
$x$, $y$, and $W = xy$.  Define ${\mathbb C}^{\times}_R$ to act
on $y$ with degree 2 and $x$ with degree 0, so that $W$ has weight two
and $-1 \in {\mathbb C}^{\times}_R$ acts trivially on $X$.
Consider the matrix factorization
\begin{displaymath}
\xymatrix{
{\cal O} \ar@/_/[d]_{P = x} \\
{\cal O} \ar@/_/[u]_{Q = y}
}
\end{displaymath}
In order to discuss the degrees of $P$ and $Q$, we must pick 
${\mathbb C}^{\times}_R$ equivariant structures on the two ${\cal O}$'s.
Since they are trivial line bundles, ${\mathbb C}^{\times}_R$
equivariant structures are in one-to-one correspondence with the integers.
Denote the equivariant structure with brackets, as ${\cal O}[n]$,
and for example take them to be
\begin{displaymath}
\xymatrix{
{\cal O}[0] \ar@/_/[d]_{P = x} \\
{\cal O}[-1] \ar@/_/[u]_{Q = y}
}
\end{displaymath}
Under the ${\mathbb C}^{\times}_R$ action, each module ${\mathbb C}[x,y]$
decomposes as
\begin{displaymath}
{\mathbb C}[x,y] \: = \: {\mathbb C}[x] \oplus y {\mathbb C}[x] \oplus
y^2 {\mathbb C}[x] \: + \: \cdots
\end{displaymath}
with a degree shift determined by the equivariant structure.  
Shifting appropriately, we find that the maps $P$, $Q$ act as follows
between sections:
\begin{displaymath}
\xymatrix{
 & 0 & 1 & 2 & 3 & 4 \\
{\cal O}[0]: & {\mathbb C}[x] \ar[dr]^x & 0 & y {\mathbb C}[x]
\ar[dr]^x & 0 & y^2 {\mathbb C}[x] \cdots \\
{\cal O}[-1]: & 0 & {\mathbb C}[x] \ar[ur]^y & 0 &
y {\mathbb C}[x] \ar[ur]^y & 0 \cdots
}
\end{displaymath}
In this fashion, we see that $x$ and $y$ each have degree one.
Note that the degrees of $x$ and $y$ are dependent upon the equivariant
structures placed on ${\cal E}_0$, ${\cal E}_1$.  For example, if
instead of ${\cal O}[-1]$ we had used ${\cal O}[-2]$, then $P$ would
have degree two and $Q$ degree zero.

The ${\mathbb C}^{\times}_R$ symmetry described above does not always exist -- 
for example, Landau-Ginzburg mirrors to non-Calabi-Yau spaces do not
admit such an R symmetry.

Given the R symmetry, we can give an alternative description of
matrix factorizations.  Specifically, define a matrix
factorization to be a pair $({\cal E}, d_{\cal E})$, where
${\cal E}$ is a ${\mathbb C}^{\times}_R$-equivariant sheaf on $X$, and
$d_{\cal E}$ is an endomorphism of ${\cal E}$ with ${\mathbb C}^{\times}_R$
weight one and such that $d_{\cal E}^2 = W 1_{\cal E}$.

Since $-1 \in {\mathbb C}^{\times}_R$ acts trivially on $X$,
${\cal E}$ splits into eigensheaves ${\cal E}_0$, ${\cal E}_1$ for the
action of that ${\mathbb Z}_2 \subset {\mathbb C}^{\times}_R$,
and $d_{\cal E}$ exchanges the two factors.  In this fashion we recover
the definition of a matrix factorization given earlier.
(Conversely, take ${\cal E} = {\cal E}_0 \oplus {\cal E}_1$ and 
proceed in the obvious fashion.)

If the action of all of ${\mathbb C}^{\times}_R$ is trivial on $X$
(as happens in an ordinary nonlinear sigma model), then necessarily
$W\equiv 0$ and ${\cal E}$ breaks into ${\mathbb Z}$-graded eigensheaves,
giving a complex of sheaves
\begin{displaymath}
\cdots \: \stackrel{d}{\longrightarrow} \: {\cal E}_{-1} \: 
\stackrel{d}{\longrightarrow} \:
{\cal E}_0 \: \stackrel{d}{\longrightarrow} \: {\cal E}_1 \: 
\stackrel{d}{\longrightarrow} \: \cdots
\end{displaymath}
in which $d^2 = 0$.  In this case, one recovers the objects of the
ordinary derived category.

In appendix~\ref{sect:tech}, we review mathematical definitions of
maps between matrix factorizations, homotopies between maps, and so forth,
and briefly compare them to corresponding physics notions in the spirit
of \cite{snowbird,al}.

\subsection{RG flow, quasi-isomorphism, and matrix factorizations}
\label{sect:quasi-rgflow}

In this paper we wish to understand D-brane probes of Landau-Ginzburg models,
which means technically constructing moduli spaces of matrix factorizations
which RG flow to D0-branes or other point-like objects.  
Therefore, we need to understand the behavior of the renormalization
group.

In the physical realization of ordinary derived categories of coherent
sheaves,
renormalization group flow
realizes localization on quasi-isomorphisms:  quasi-isomorphic
complexes define boundary states in the same universality class of
renormalization group flow.

Here, one would expect that renormalization group flow should similarly
correspond to inversion of quasi-isomorphisms.  However, since our
`complexes' are no longer complexes (successive maps do not compose to zero),
we first need to find a suitable definition of `quasi-isomorphism', that relates different matrix factorizations that are in the
same universality class.

There is another problem related to RG flow: in many interesting examples a Landau-
Ginzburg model RG flows to a non-linear sigma model on some target $Y$. In these examples, we seek a functor from the
category of matrix factorizations to the derived category of coherent sheaves on $Y$, specifying where the D-branes flow.

\subsubsection{Quasi-isomorphisms between matrix factorizations}
\label{sect:quasiiso}

For matrix factorizations, the standard definition of
quasi-isomorphism (a map between complexes that induces an isomorphism
on homology) does not make sense, since unless $W$ vanishes identically our D-branes are not
honest complexes, so we cannot talk about their homology. However, finding a suitable notion of quasi-isomorphism
is important for physics, as it will surely generalize
criteria for two matrix factorizations to be in the same universality class. 
Fortunately, an appropriate definition has appeared in the mathematics 
literature ({\it e.g.} \cite{orlov11}). 
We will discuss it in this section, focusing on a few basic examples. 
More details are given in appendix~\ref{sect:tech}.

Firstly, recall that for ordinary complexes, a map $f$ is a quasi-isomorphism
if and only if the cone $C(f)$ is acyclic \cite[corollary 1.5.4]{weibel}, 
{\it i.e.} $C(f)$ has no homology. If $f$ is instead a map between matrix factorizations, then we can still define the cone $C(f)$ in exactly the same way, but now $C(f)$ is also a matrix factorization so it makes no sense to ask if $C(f)$ is acyclic. However, if we can come up with a definition of an `acyclic' matrix factorization, then we can get a definition of `quasi-isomorphism' for free, because we can declare that $f$ is a quasi-isomorphism iff $C(f)$ is acyclic.

 Suppose ${\cal E}, {\cal F}, {\cal G}, \dotsc$ are ordinary complexes, and that we have an exact sequence of complexes
$$ 0 \rightarrow {\cal E} \stackrel{f}{\rightarrow} {\cal F}  \stackrel{g}{\rightarrow} {\cal G} \rightarrow \dotsb$$
{\it i.e.} $f, g, \dotsc$ are chain maps, and in each degree they define an exact sequence of sheaves. Then the iterated cone on the maps $f, g, \dotsc$ defines a single complex, and this complex will be acyclic. Conversely, every acyclic complex arises in this way, because an acyclic complex is precisely an exact sequence of sheaves, and we can view each sheaf as a complex of length one.

Now suppose that ${\cal E},{\cal F},{\cal G}, \dotsc$ are instead matrix factorizations. Then it still makes sense to ask if the maps $f, g, \dotsc$ define an exact sequence, because we just ask if we have an exact sequence of sheaves in each degree. So we can \emph{define} a matrix factorization to be acyclic iff it is (homotopy equivalent to) an iterated cone over an exact sequence of matrix factorizations. Then as we said above, we define a map $f$ to be a quasi-isomorphism iff $C(f)$ is acyclic.

Let us work through some simple examples. Suppose we have a 
Landau-Ginzburg model $(X,W)$ where $X$ is the total space of a line bundle
$$\pi: {\cal L} \rightarrow B$$
and the superpotential is $W = p \pi^* s$, where
$p \in \Gamma_X( \pi^*{\cal L} )$ is a fiber coordinate and $s \in \Gamma_B( {\cal L}^*)$.  It is 
possible to put a $\mathbb{C}^\times_R$ R-charge on this model, 
but we won't worry about this for the moment.

On $X$ we have a sheafy matrix factorization
\begin{displaymath}
\xymatrix{
{\cal O}_B \ar@/_/[d] \\
0 \ar@/_/[u]
}
\end{displaymath}
(where 
 ${\cal O}_B$ is supported on
the zero section of ${\cal L}$),
which we shall often denote merely ${\cal O}_B$. We claim that this sheafy matrix factorization is  quasi-isomorphic to the
traditional matrix factorization
%\begin{displaymath}
%{\cal O}_X \: \stackrel{s}{\longrightarrow} \:
%{\cal L}^* \: \stackrel{p}{\longrightarrow} \: {\cal O}_X
%\end{displaymath}
\begin{displaymath}
\xymatrix{
{\cal O}_X \ar@/_/[d]_{s} \\
\pi^* {\cal L}^* \ar@/_/[u]_p
}
\end{displaymath}
To see this, consider the cone over the map
\begin{equation}\label{egofquasiisomorphism1}
\xymatrix{
{\cal O}_X \ar[r] \ar@/_/[d]_s 
 & {\cal O}_B \\
\pi^* {\cal L}^* \ar@/_/[u]_p
}
\end{equation}
This is homotopy-equivalent to the iterated cone over the
exact sequence
\begin{equation}\label{exactsequenceofMFs1}
\xymatrix{
{\cal L}^* \ar[r]^p \ar@/_/[d]_W 
& {\cal O}_X \ar[r] \ar@/_/[d]_s 
 & {\cal O}_B \\
\pi^* {\cal L}^* \ar[r]^1 \ar@/_/[u]_1
& \pi^* {\cal L}^* \ar@/_/[u]_p
}
\end{equation}
since the matrix factorization
\begin{displaymath}
\xymatrix{
\pi^* {\cal L}^* \ar@/_/[d]_W \\
\pi^* {\cal L}^* \ar@/_/[u]_1
}
\end{displaymath}
is (easily checked to be) contractible. So by definition, the map (\ref{egofquasiisomorphism1}) is a quasi-isomorphism (see appendix~\ref{sect:tech} for more details). Thus, we propose that renormalization
group flow identifies ${\cal O}_B$ and the traditional matrix factorization
%\begin{displaymath}
%{\cal O}_X \: \stackrel{s}{\longrightarrow} \:
%{\cal L}^* \: \stackrel{p}{\longrightarrow} \: {\cal O}_X
%\end{displaymath}
\begin{displaymath}
\xymatrix{
{\cal O}_X \ar@/_/[d]_{s} \\
\pi^* {\cal L}^* \ar@/_/[u]_p
}
\end{displaymath}
(Note that we have not yet discussed endpoints of RG flow, which we will
cover in the next section; we are merely
proposing that these two matrix factorizations flow to the same endpoint,
whatever that endpoint might be.)
Physically, we think of the contractible matrix factorization
\begin{displaymath}
\xymatrix{
\pi^* {\cal L}^* \ar@/_/[d]_W \\
\pi^* {\cal L}^* \ar@/_/[u]_1
}
\end{displaymath}
as merely providing spectators, that drop out of renormalization group flow
but illuminate its direction.

A closely related example is as follows. 
In the Landau-Ginzburg model
above, we have another sheafy matrix factorization ${\cal O}_{\{\pi^*s=0\}} 
$ (in general when we list a single sheaf as a sheafy matrix factorization,
we mean to indicate that the second sheaf vanishes identically, as in the
notation of the previous example). It is
quasi-isomorphic to the traditional matrix factorization
\begin{displaymath}
\xymatrix{
{\cal O}_X \ar@/_/[d]_{p} \\
\pi^* {\cal L} \ar@/_/[u]_{s}
}
\end{displaymath}
by virtue of the short exact sequence of matrix factorizations
\begin{equation}\label{exactsequenceofMFs2}
\xymatrix{
\pi^* {\cal L} \ar@/_/[d]_W \ar[r]^s &
{\cal O}_X \ar@/_/[d]_p \ar[r] &
{\cal O}_{\{\pi^*s=0\}} \\
\pi^* {\cal L} \ar@/_/[u]_1 \ar[r]^1 &
\pi^* {\cal L} \ar@/_/[u]_s 
}
\end{equation}

Next, let us consider the sheafy matrix factorization ${\cal O}_{\{s=0\}}$, 
supported on the locus $\{s=0\} \subset B \subset X$.  Let us find a quasi-isomorphic traditional matrix factorization.
Here, the support has codimension two, so we will have to work a bit
harder than in the last examples.  Appendix~\ref{sect:tech} outlines
a general procedure for computing resolutions of sheafy matrix factorizations
supported on $\{W=0\}$;
we shall explicitly describe some of the details in this example.

Let us begin with the Koszul resolution of the sheaf ${\cal O}_{\{s=0\}}$
on $X$:
\begin{displaymath}
0 \: \longrightarrow \: {\cal O}_X \: \stackrel{[-p, s]^T}{\longrightarrow} \:
\pi^* {\cal L} \oplus \pi^* {\cal L}^* \: \stackrel{[s, p]}{\longrightarrow} \:
{\cal O}_X \: \stackrel{1}{\longrightarrow} \: {\cal O}_{\{s=0\}}
\: \longrightarrow \: 0
\end{displaymath}
We can fold the Koszul resolution into a two-term complex:
\begin{displaymath}
\xymatrix{
{\cal O}_X \oplus {\cal O}_X \ar@/_/[d]_a \\
\pi^* {\cal L} \oplus \pi^* {\cal L}^* \ar@/_/[u]_b
}
\end{displaymath}
where
\begin{displaymath}
a \: = \: \left[ \begin{array}{cc}
-p & 0 \\
s & 0 \end{array} \right], \: \: \:
b \: = \: \left[ \begin{array}{cc}
0 & 0 \\
s & p \end{array} \right]
\end{displaymath}
However, this is not itself a matrix factorization, as
$ab = 0 = ba$.  To get a matrix factorization, we must add
`backwards' maps so as to get the composition to be proportional to $W$.
It is straightforward to check that if we replace $a$, $b$ above by
\begin{displaymath}
a' \: = \: \left[ \begin{array}{cc}
-p & p/2 \\
 s & s/2 \end{array} \right], \: \: \:
b' \: = \: \left[ \begin{array}{cc}
-s/2 & p/2 \\
s & p \end{array} \right]
\end{displaymath}
then $a'b' = b'a' = W \cdot {\rm Id}$, as needed for a matrix factorization.

To show that ${\cal O}_{\{s=0\}}$ 
is quasi-isomorphic to the
matrix factorization above, what we have to do is essentially tensor together 
the two exact sequences (\ref{exactsequenceofMFs1}) and (\ref{exactsequenceofMFs2})
 (and throw in some factors of $\frac12$). 
This produces an exact sequence with the shape
\begin{displaymath}
\xymatrix{
{\cal O}_X \oplus {\cal O}_X\ar@/_/[d] \ar[r] &
\pi^* {\cal L} \oplus {\cal O}_X \oplus \pi^* {\cal L}^* \oplus {\cal O}_X 
\ar@/_/[d] \ar[r] &
{\cal O}_X \oplus {\cal O}_X \ar@/_/[d]_{a'} \ar[r]^{[0, 1]} &
{\cal O}_{\{s=0\}} \\
{\cal O}_X  \oplus {\cal O}_X\ar@/_/[u] \ar[r] &
\pi^* {\cal L} \oplus {\cal O}_X \oplus {\cal O}_X \oplus \pi^* {\cal L}^*
\ar@/_/[u] \ar[r]&
\pi^* {\cal L} \oplus \pi^* {\cal L}^* \ar@/_/[u]_{b'}
}
\end{displaymath}
in which the first two matrix factorizations are contractible.

 In fact it is always possible to replace a sheafy matrix factorization by a quasi-isomorphic traditional matrix factorization (just as any sheaf can be resolved by vector bundles). However, allowing matrix factorizations to be sheafy gives us much more flexibility when working with them, as well as making certain aspects of their behavior more explicit.

As an aside, if we only work with traditional matrix factorizations, and additionally assume that $X$ is affine, then it can be shown that quasi-isomorphism reduces to homotopy-equivalence
(essentially from the fact that every short exact sequence of vector
bundles splits).  For this reason, 
references such as \cite{orlov03} build categories
of matrix factorizations in which one merely quotients out homotopy
equivalences (see {\it e.g.} the category $DB$ defined in
\cite{orlov03}), and do not need to localize further on quasi-isomorphisms.

\subsubsection{Functors from matrix factorizations in LG to sheaves in NLSM}
\label{sect:RG:functor}

Suppose, as a slight generalization of the above examples, that we have a Landau-Ginzburg model $(X,W)$ where $X$ is the total space of a {\it vector} bundle
$$\pi: X \rightarrow B$$
 and the superpotential is $W = p \pi^* s$, where
$p \in \Gamma_X(\pi^* X)$ is a fiber coordinate and $s \in \Gamma_B( X^*)$. 
Let us further assume that
the locus $\{s = 0\} \subset B$ is smooth, so that the critical locus of $W$ is
\begin{eqnarray*}
Y &:=& \{ dW = 0 \} \subset X \\
&=& \{s = 0 \} \subset B
\end{eqnarray*}
We will also define an R-charge $\mathbb{C}^\times_R$ so that it fixes $B$, 
and acts on the fibers with weight two, so $p$ has weight two, 
and $\pi^*s$ has weight zero. 

Under RG flow, this model should flow to the non-linear sigma model on $Y$. 
Consequently, there should be a functor from matrix factorizations on $X$ to 
sheaves on $Y$ which is an equivalence of categories. This equivalence, 
which is a family version of classical Kn\"orrer periodicity \cite{knorrer}, 
is well-studied in the mathematics literature 
({\it e.g.} \cite{isik10, orlovKP, shipman}), and also appeared in \cite{hhp}. 
For a given sheafy matrix factorization $({\cal E}_0, {\cal E}_1,
P, Q)$, the functor consists of the following steps:
\begin{enumerate}
\item Replace the sheafy matrix factorization by a quasi-isomorphic
traditional
matrix factorization built from vector bundles.
\item Restrict to the locus $\{\pi^*s=0\}$, which is a vector bundle over $Y$. 
Now we have a matrix factorization of $W=0$.\footnote{One should think of 
these first two steps together as `derived restriction.'}
\item Push down to $Y$.  
\item Now that we are on $Y$, the ${\mathbb C}^{\times}_R$
acts trivially, so our matrix factorization can be unrolled into a
complex.  It will have finite homology, and so lives in $D^b(Y)$.
\end{enumerate}
If we are interested in the homology of the complex just obtained,
we can take the homology of the matrix factorization obtained at the
second step, which is an ${\mathbb C}^{\times}_R$-equivariant sheaf,
and  push it down to get a graded sheaf on the critical locus.  
(Since the pushdown functor is exact, it commutes with the operation of
taking homology.)
 
We will present several examples of the results of this functor,
and discuss their interpretation, to give consistency checks that this
mathematical description of renormalization group flow is correct.

Let us specialize to $X$ being the total space
of a line bundle $\pi: {\cal L} \rightarrow B$, and consider the sheafy
matrix factorization given by the sheaf ${\cal O}_B$, which we discussed in the previous section. We showed that ${\cal O}_B$ 
is quasi-isomorphic to the traditional
matrix factorization
\begin{displaymath}
\xymatrix{
{\cal O}_X \ar@/_/[d]_s \\
\pi^* {\cal L}^*[-1] \ar@/_/[u]_p
}
\end{displaymath}
Note that we are now taking account of the R-symmetry -- the pulled-up 
line bundle ${\cal L}^*$ comes with a natural trivial 
$\mathbb{C}^\times_R$-equivariant structure\footnote{Because over any single 
orbit the bundle is canonically trivial.}, and we are writing ${\cal L}^*[-1]$ 
to denote the twist of this equivariant structure by the $-1$ character of 
${\mathbb C}^\times_R$.  In the previous section we neglected the R-symmetry, 
but including it does not change the argument. 

When we restrict to $\{\pi^*s=0\}$, the map $s$ becomes the zero map, so this matrix factorization becomes a two-term complex of bundles with homology given by
${\cal O}_Y$.  Pushing down gives the sheaf ${\cal O}_Y$
in $D^b(Y)$.

For a second example, start with the sheafy matrix
factorization ${\cal O}|_{\{\pi^*s=0\}}$.  
This is equivalent 
to the traditional
matrix factorization
\begin{displaymath}
\xymatrix{
{\cal O}_X \ar@/_/[d]_p \\
\pi^* {\cal L}[1] \ar@/_/[u]_s
}
\end{displaymath}
Restricting to $\{\pi^*s=0\}$ and pushing forward, we get the object
${\cal L}_Y[1]$ in $D^b(Y)$.

Now suppose we start with the sheafy matrix factorization
${\cal O}_Y $, the skyscraper sheaf along the critical locus.  This is quasi-isomorphic to the traditional matrix factorization
\begin{displaymath}
\xymatrix{
{\cal O}_X \oplus {\cal O}_X \ar@/_/[d]_{a'} \\
\pi^* {\cal L}[1]
 \oplus \pi^* {\cal L}^*[-1] \ar@/_/[u]_{b'}
}
\end{displaymath}
where
\begin{displaymath}
a' \: = \: \left[ \begin{array}{cc}
-p & p/2 \\
 s & s/2 \end{array} \right], \: \: \:
b' \: = \: \left[ \begin{array}{cc}
-s/2 & p/2 \\
 s & p \end{array} \right]
\end{displaymath}
When we restrict to $\{\pi^*s=0\}$ this becomes a complex, which we can write as
\begin{displaymath}
0 \: \longrightarrow \: \pi^* {\cal L}^*[-1] \: \stackrel{[p,0]^T}{\longrightarrow} \:
{\cal O}_X \oplus {\cal O}_X \: \stackrel{[p,0]}{\longrightarrow} \:
\pi^* {\cal L}[1] \: \longrightarrow \: 0
\end{displaymath}
(we have changed basis in ${\cal O}_X \oplus {\cal O}_X$). 
Taking homology and pushing-down, we get
$${\cal O}_Y \oplus {\cal L}_Y[1]$$
{\it i.e.}  the direct sum of the previous two examples. Since this functor is an equivalence, this means that ${\cal O}_Y$ must be quasi-isomorphic, as a sheafy matrix factorization, to the direct sum of ${\cal O}_B$ and ${\cal O}_{\{\pi^*s=0\}}$.  In fact, if you replace all three by quasi-isomorphic traditional 
matrix factorizations in the way that we have described, it is easy to find an 
explicit isomorphism between the first one and the direct sum of the 
second two.

This last example is particularly curious, in that we start with D-branes
on the critical locus $Y=\{dW=0\}$, which RG flow to a \emph{different} set of branes on
$Y$, whereas one might naively have thought that RG flow would leave
invariant anything supported only on the critical locus. However, as pointed out in
\cite{kapli}, even a massive Landau-Ginzburg 
model can have massless open string states,
so we must be careful about the `massive' directions in the noncompact
Landau-Ginzburg model above.

As a consistency check, let us compare Ext groups.
Using the result from appendix~\ref{sect:tech}, Ext groups (open string
states) between the sheafy matrix factorization above and itself are given by
\begin{displaymath}
{\rm Ext}^*_{MF}({\cal O}_Y, 
{\cal O}_Y) \: = \:
H^*(Y, \textstyle\bigwedge^* (N_{Y/X}[1]) )
\end{displaymath}
(we are using the fact that 
since $dW=0$ on this locus, the differentials in the
spectral sequence are all trivial).
By contrast, Ext groups in the category of sheaves on the critical
locus are given by
\begin{displaymath}
{\rm Ext}^*_Y({\cal O}_Y, {\cal O}_Y) \: = \:
H^*(Y, {\cal O}_Y)
\end{displaymath}
If the sheafy matrix factorization ${\cal O}_Y$
RG flowed to the sheaf ${\cal O}_Y$ on the critical locus,
then the open string states in the B model (the Ext groups) would be preserved,
but we see that is not the case.  There are more Ext group elements in the
category of matrix factorizations (arising ultimately from the fact that there
are extra $\theta$ fields in the worldsheet B twist),
Thus, the sheafy matrix factorization
${\cal O}_Y$ cannot RG flow to the sheaf ${\cal O}_Y$
on the critical locus.
On the other hand, our claim that it flows to 
${\cal O}_Y \oplus {\cal L}_Y[1]$ 
is entirely consistent with this calculation, 
because 
\begin{eqnarray*}
 {\rm Ext}^*_Y( \,{\cal O}_Y \oplus {\cal L}_Y[1], \,{\cal O}_Y \oplus {\cal L}_Y[1]) & = & H^*(Y,\: {\cal O}_Y\oplus {\cal L}_Y[1]\oplus 
{\cal L}^*_Y[-1] \oplus {\cal O}_Y)\\
&=&H^*(Y,\: \textstyle\bigwedge^*( {\cal L}_Y[1]\oplus {\cal L}^*_Y[-1]))
\end{eqnarray*}
and $N_{Y/X} = {\cal L}_Y\oplus {\cal L}^*_Y[-2]$.

\subsection{Point-like objects in matrix factorizations}
\label{sect:pointlike}

This paper is concerned with using sheafy matrix
factorizations to provide D-brane probes of SCFT's obtained from certain
Landau-Ginzburg models describing branched double covers and 
(sometimes) noncommutative resolutions thereof.
In nonlinear sigma models, D-brane probe moduli spaces are moduli spaces
of D0-branes, which on a space have pointlike support.
In the present case, because we are working in Landau-Ginzburg models,
we have to be slightly more careful.

Since all of the Landau-Ginzburg models we consider will RG flow
to nonlinear sigma models on either smooth manifolds or noncommutative
resolutions, morally we want our `point-like' matrix factorizations
to 
RG flow to D0-branes in the IR nonlinear sigma model.  

If we only ever worked with Landau-Ginzburg models that RG flowed
to nonlinear sigma models on smooth manifolds, this definition
would suffice;
however, we are also interested in Landau-Ginzburg models that RG flow
to CFT's defined by noncommutative resolutions of singular spaces, and so
we need a different notion of point-like, one that is well-defined in
greater generality.

To begin to grasp the pertinent issues, let us walk through some
special cases.
One example of a matrix factorization that should be 
considered point-like
is implicit earlier in this section:
a skyscraper sheaf supported at a point
is a trivial example of a matrix factorization (trivial, as since there
is no Warner problem to contend with, no second brane or tachyon maps are
required).

However, that is not the only possibility.  For example, a skyscraper sheaf
could be quasi-isomorphic to a nontrivial matrix factorization, involving
sheaves not necessarily supported over the critical locus of the 
superpotential.  In such a case, the point-like nature of the matrix
factorization would not be immediately apparent. 

In this section we will describe two procedures for checking, to at
least a partial degree, whether an object is `pointlike'.
The first method is as follows:  we can define the (set-theoretic)
support of a matrix factorization to be the smallest locus $S \subset X$
such that the matrix factorization is contractible on $X-S$.

Here is a simple example.  Consider the Landau-Ginzburg model on
$X = {\mathbb C}^2$, with $W = xp$.  Consider the matrix factorization
\begin{displaymath}
\xymatrix{
{\cal O} \ar@/_/[d]_x \\
{\cal O} \ar@/_/[u]_p
}
\end{displaymath}

First, let us show that its support lies within the locus $\{p=0\}$,
by showing that it is contractible elsewhere.  To do this, we need
to find maps $s, t: {\cal O} \rightarrow {\cal O}$ such that
\begin{displaymath}
1 \: = \: ps \: + \: tx
\end{displaymath}
This is easily solved by taking $t=0$ and $s = p^{-1}$, hence,
the matrix factorization is contractible if we restrict to the locus
$\{ p \neq 0 \}$. Similarly, it can also be shown to be contractible
on the locus $\{ x \neq 0 \}$, and hence the set-theoretic support of
this matrix factorization consists of the point
$\{x=p=0\}$.

As an aside, notice that there is a general result here: for an arbitrary superpotential, the matrix factorization
\begin{displaymath}
\xymatrix{
{\cal O} \ar@/_/[d]_{\lambda} \\
{\cal O} \ar@/_/[u]_{W \lambda^{-1}}
}
\end{displaymath}
is contractible for any never-vanishing function $\lambda$. We're using the special case $W=xp$, and $\lambda = x$ (or $p$).

However, this criterion of set-theoretic support is not completely adequate.
The reason is that there may be multiple coincident skyscraper sheaves
at a single point, in which case, the set-theoretic support would still
be a single point, but sheaf-theoretically we would be describing
multiple coincident points. 

Therefore, we supplement the notion of set-theoretic support with a second
test, utilizing homological algebra.  We say that a matrix factorization ${\cal E}$
is \emph{homologically pointlike} if it has the same Ext groups with itself
as the skyscraper sheaf of point in the ordinary derived category, 
{\it i.e.} if there is some $n$ such that
$$\dim {\rm Ext}^k_{MF}({\cal E}, {\cal E}) =  {n \choose k }$$
for each $k$ (since these are the dimensions of the self-Ext groups of a skyscraper sheaf on a point of an $n$-dimensional manifold).

Let's apply this homological test in our previous simple example where we had  $X=\mathbb{C}^2$ with coordinates $x$ and $p$, and superpotential $W=xp$. For more precision, let's recall that we can equip the model with a $\mathbb{C}^\times_R$ R-charge by letting $p$ have weight 2, and $x$ have weight 0. Then we have a matrix factorization ${\cal E}$ given by
\begin{displaymath}
\xymatrix{
{\cal O} \ar@/_/[d]_x \\
{\cal O}[-1] \ar@/_/[u]_p
}
\end{displaymath}
 We follow the procedure and notation from appendix~\ref{sect:tech}
to compute Ext$^i_{\rm MF}({\cal E},{\cal E})$. We have
\begin{displaymath}
{\cal H}om_0 \: = \: {\cal H}om({\cal O},{\cal O}) \oplus {\cal H}om({\cal O}[-1],
{\cal O}[-1]) \: = \: {\cal O} \oplus {\cal O}
\end{displaymath}
and
\begin{displaymath}
{\cal H}om_1 \: = \: {\cal H}om({\cal O},{\cal O}[-1]) \oplus {\cal H}om({\cal O}[-1],
{\cal O}) \: = \: {\cal O}[-1] \oplus {\cal O}[1]
\end{displaymath}
with maps
\begin{displaymath}
P \: = \: \left[ \begin{array}{cc}
x & -x \\
-p & p \end{array} \right], \: \: \:
Q \: = \: \left[ \begin{array}{cc}
p & x \\
p & x \end{array} \right]
\end{displaymath}

Next we take global sections of these local Hom bundles. Since we have a 
$\mathbb{C}^\times_R$ R-symmetry, the global sections split up into 
integer-graded pieces, and we get an honest chain complex of vector spaces:
\begin{displaymath}
\xymatrix{
-1 & 0 & 1 & 2 & 3  \\
0& {\mathbb C}[x]\oplus {\mathbb C}[x] 
\ar[dr]^(.6){\; \left[\begin{smallmatrix}  x&-x \\
-p&p \end{smallmatrix}\right]}
 & 0 & p {\mathbb C}[x]\oplus p{\mathbb C}[x]
\ar[dr]^(.6){\; \left[\begin{smallmatrix} x & -x \\
-p & p \end{smallmatrix}\right]}
 & 0  \cdots \\
 {\mathbb C}[x] \ar[ur]^(.4){ \left[\begin{smallmatrix}  x \\
x \end{smallmatrix}\right]} & 0 &
  {\mathbb C}[x]\oplus p{\mathbb C}[x]
 \ar[ur]^(.4){ \left[\begin{smallmatrix} p & x \\
p & x \end{smallmatrix}\right]}
 & 0 &
p^2 {\mathbb C}[x]\oplus p{\mathbb C}[x]   \cdots
}
\end{displaymath}
Taking homology of this chain complex, we compute the result:
\begin{displaymath}
{\rm Ext}^0_{\rm MF}({\cal E},{\cal E}) \: = \: \mathbb{C}, \: \: \:
{\rm Ext}^i_{\rm MF}({\cal E},{\cal E}) \: = \: 0 \mbox{ for }i \neq 0.
\end{displaymath}
This is consistent with the self-Ext's of a skyscraper sheaf of a point living in a 0-dimensional manifold, 
{\it i.e.} an isolated point not embedded in a higher-dimensional space.

Notice that these calculations are consistent with our calculations of RG flow in the previous section. If we let $B$ be the 1-dimensional space $\mathbb{C}$ with coordinate $x$, then $X$ is theof total space of the trivial line-bundle
$$\pi: X=\mathbb{C}^2 \to \mathbb{C} = B $$
This is a very simple case of the more vector bundle general model considered previously, the section $s$ in this case is just the function $x$. Therefore, this model should RG flow to the sigma model whose target space is just the single point $Y=\{x=p=0\}$. By our previous arguments, the matrix factorization ${\cal E}$, which is quasi-isomorphic to the sheafy matrix factorization ${\cal O}_B$, must RG flow to the structure sheaf ${\cal O}_Y$.  So it is fortunate that the Ext groups agree.

We have given two criteria now for `point-likeness', 
but we will not claim that our criteria are either complete or 
fool-proof.  For example, on Fano varieties there is a slightly stronger notion of point-like which allows one to single out the actual skyscraper sheaves of points in purely categorical terms \cite{bondal-orlov-reconstruction}.  In the Calabi-Yau case, which is our primary interest, this stronger criterion is not applicable, and there may be several classes of point-like objects, reflecting the possibility of several Calabi-Yau manifolds having equivalent derived categories.

\section{D-brane probes of smooth branched double covers}
\label{sect:smooth}

In this section, we will discuss D-brane probes of Landau-Ginzburg
models that are believed to
flow in the IR to nonlinear sigma models on smooth branched
double covers, following the pattern described in \cite{cdhps}.

We begin by reviewing basic aspects of D-brane probes of Landau-Ginzburg
models over nontrivial spaces, realized via sheafy matrix factorizations,
and also general aspects of the branched double cover realization.
We then turn to particular examples.  Most examples of interest
are related (via K\"ahler moduli) to complete intersections of quadric
hypersurfaces, for which the one quadric special case forms the prototype
for behavior in fibers.  Our analysis of examples involving multiple
quadrics is primarily local in nature, but at the end of this
section, we discuss global gluing issues, and how topologically
nontrivial $B$ fields sometimes arise.  

The results in this section are not particularly surprising, in that
they merely recover the results of \cite{cdhps} (namely, that certain
Landau-Ginzburg models will flow in the IR to nonlinear sigma models
on branched double covers), albeit via novel methods.  In the next section,
we study D-brane probes of non-geometric `nc resolutions'
also encountered in \cite{cdhps}, which gives genuinely new results and
insight into the nature of those conformal field theories.

The mathematics outlined in this section is already present in \cite{nick1},
and
we refer the interested reader there for greater detail.
Our contribution here is the application of that mathematics 
via sheafy matrix factorizations to
physics questions of D-brane probes.

\subsection{D-brane probes of standard vector bundle models}

A D-brane probe of a Landau-Ginzburg model is, at least morally, a 
D0-brane propagating inside the Landau-Ginzburg model. In this paper, we are primarily interested in Landau-Ginzburg models
that RG flow to branched double covers, but as a warmup exercise,
let us briefly consider a more standard case.

Consider (as we have previously) a Landau-Ginzburg model defined on the total space of a vector bundle
\begin{displaymath}
X \: = \: {\rm Tot}\left( {\cal V} \: \stackrel{\pi}{\longrightarrow}
\: B \right)
\end{displaymath}
with superpotential $W = p \pi^* s$, where $p$ is a fiber coordinate and $s$ is a transverse section of the dual bundle ${\cal V}^*$ over the base $B$. In the IR, this model flows to a nonlinear sigma model on the critical locus
$$Y:= \{s =0\} \subset B $$
 Because of the RG endpoint, we should be able to find $Y$ as a moduli space of `point-like' branes in the original model. Let's verify this.

Pick a point $y\in Y$, and consider the fiber ${\cal V}_y$ of the bundle over 
this point. This is  a submanifold of $X$, and it lies within the locus 
$\{W =0\}$. 
Hence ${\cal O}_{{\cal V}_y}$ is a well-defined sheafy matrix 
factorization\footnote{
Alternatively, we could consider a sheafy matrix factorization defined
by a skyscraper sheaf.  Since there is a potential term $| dW |^2$, and we
wish to describe zero-energy motions, we should restrict to skyscraper
sheaves supported on the locus $\{ dW = 0 \}$.  This gives another way of
thinking about moduli spaces of D0-branes in this model, and also reflects
the fact that the bulk theory localizes on maps into the
critical locus $\{ dW = 0 \}$, since $\delta \theta \propto d W$.  
That said, this example need not RG flow to a single 
skyscraper sheaf on the critical locus, and should
not be interpreted as a single D0-brane, but in general will RG flow to a 
collection of skyscraper sheaves moving in sync, and so can also be used to
map out the D0-brane moduli space.
}.

We claim that this sheafy matrix factorization satisifies both the criteria for 
point-likeness given in section~\ref{sect:pointlike}. 
To see this, we use the general result from the end of 
appendix~\ref{app:Extgroups} to conclude that
$${\rm R}{\cal H}om_{MF} ( {\cal O}_{{\cal V}_y}, {\cal O}_{{\cal V}_y}) 
\: = \: \textstyle\bigwedge(N_{{\cal V}_y/X}[1]) $$
where the right-hand-side carries the differential `contract with $dW$'. 

This answer is only sensitive to a first-order neighborhood of ${\cal V}_y$, so we are free to reduce to the case that $B=\mathbb{C}^n$ with origin at $y$, the bundle ${\cal V}$ is trivial (of rank $k$ say), and the section $s$ is
$$ s = (x_1,\cdots,x_k) \in \Gamma (B, {\cal O}^k) $$
(the $x_i$ are coordinates on $B$). Then $W$ is given by the degenerate quadratic form
$$W \: = \: x_1 p_1 \: + \: \cdots \: + \: x_k p_k $$
and ${\cal V}_y$ is the isotropic subspace $\{x_1=\cdots=x_n=0\}$. 
Then it is a straight-forward explicit calculation (which we give in 
appendix~\ref{app:pointlike}) to show that 
$${\rm Ext}^*_{MF} ( {\cal O}_{{\cal V}_y}, {\cal O}_{{\cal V}_y}) \:
=  \: \textstyle\bigwedge( T_y Y[1] ) $$
(which verifies homological point-likeness) and also that that the set-theoretic support of this matrix factorization is just the point $y$. 

Obviously, we can identify the set of all such branes with the set of points in $Y$. Furthermore, the above calculation tells us that the deformation theory of these branes matches exactly with the deformations of the corresponding points in $Y$, so the moduli space of these branes really is the space $Y$. 

For completeness, we should discuss the effect of RG flow on these branes. As a first guess, one might expect that ${\cal O}_{{\cal V}_y}$ flows to the sky-scraper sheaf ${\cal O}_y \in D^b(Y)$. However, careful application of the recipe from section~\ref{sect:RG:functor} shows that in fact it flows to the object
$$({\rm det} {\cal V})|_y\, [{\rm rk}({\cal V})] $$
Fortunately these objects have the same moduli space as the sheaves ${\cal O}_y$, since they are related by the operation `twist by the line-bundle $({\rm det} {\cal V})$, then shift by ${\rm rk}({\cal V})$'. This is an autoequivalence of $D^b(Y)$, so in particular it preserves moduli spaces.

Here's a quick consistency check on the above claim: if we stick all these branes together into a family, we get the sheafy matrix factorization ${\cal O}_{{\cal V}_Y}$. It follows that this should RG-flow to the shifted line-bundle $({\rm det} {\cal V}) [{\rm rk}({\cal V})] \in D^b(Y)$. We verified this in 
section~\ref{sect:RG:functor} in the case that ${\cal V}$ has rank 1.

\subsection{Branched double covers}

In the previous section, we studied some simple examples of D-brane
probes of Landau-Ginzburg models that RG flowed to ordinary NLSM's.
In this section we will study some slightly more complicated examples,
involving Landau-Ginzburg models on bundles over gerbes,
that RG flow to branched double covers and related objects.  
The structure of such
Landau-Ginzburg models was discussed previously in \cite{cdhps,kuz2}.
Briefly, these appear in GLSM's describing complete intersections of
quadric hypersurfaces.  For example, for a complete intersection of
$k$ quadric hypersurfaces in ${\mathbb P}^n$, there is a GLSM which,
at large positive radius, flows to an intermediate Landau-Ginzburg model
on  
\begin{equation}    \label{totspace:first}
{\rm Tot}\left( {\cal O}(-2)^k \: \longrightarrow \:
{\mathbb P}^n \right)
\end{equation}
with superpotential of the form
\begin{displaymath}
W \: = \: \sum_{a=1}^k p_a Q_a(\phi) \: = \: \sum_{i,j=1}^{n+1}
\phi_i \phi_j A^{ij}(p)
\end{displaymath}
which further flows to a NLSM on a complete intersection of $k$ quadrics.
At large negative radius, the intermediate Landau-Ginzburg model is
on\footnote{
The notation ${\cal O}(-\tfrac12)$ indicates a line bundle over 
${\mathbb P}^n_{[2,\cdots,2]}$ (an example of a ``${\mathbb Z}_2$ gerbe'')
defined by a nontrivial equivariant structure
under a ${\mathbb Z}_2$ that acts trivially on the base space.
The point is that the first total space, in~(\ref{totspace:first}),
is birational to the second total space, in~(\ref{totspace:z2gerbe}).
See for example \cite{hhpsa,cdhps,alg22} 
for further information on this notation.
}
\begin{equation}    \label{totspace:z2gerbe}
{\rm Tot}\left( {\cal O}(-\tfrac12)^{n+1} \: \longrightarrow \:
{\mathbb P}^{k-1}_{[2,2,\cdots,2]} \right).
\end{equation}
If $n+1$ is even, then as discussed in
section~\ref{sect:glsm:nc} and \cite{cdhps},
flows to a (possibly noncommutative resolution of a) branched double cover of
${\mathbb P}^{k-1}$.  If $n+1$ is odd, then this flows to a (possibly 
noncommutative resolution of a) single copy of ${\mathbb P}^{k-1}$, with
a hypersurface of ${\mathbb Z}_2$ orbifolds. In this paper we will mostly 
focus on cases where $n+1$ is even, 
but we will make some mention of the odd case.

In this section, we will study simple cases in which there is no noncommutative
resolution, in which the branched double cover (or single cover with
${\mathbb Z}_2$ orbifolds) is smooth.
We will study D-brane probes of such
Landau-Ginzburg models.  We will study D-brane probes of
noncommutative resolutions obtained similarly, in the next section.

\subsubsection{One quadric}   \label{sect:onequadric}

Let us begin with simple examples involving just a single quadric
hypersurface, {\it i.e.} $k=1$.  In this case, the Landau-Ginzburg model
we are interested in lives on
\begin{displaymath}
{\rm Tot}\left( \oplus_1^{n+1} {\cal O}(-\tfrac12) \: \longrightarrow \:
[ {\rm pt}/{\mathbb Z}_2] \right) \: = \: \left[
{\mathbb C}^{n+1}/{\mathbb Z}_2 \right]
\end{displaymath} 
where the ${\mathbb Z}_2$ acts by sign flips.
This will be the prototype for fibers in cases we will examine later.
For a generic quadric, there is a single isolated critical point,
at the origin of $V \equiv {\mathbb C}^{n+1}$, and all normal directions are
massive.  In the language of GLSM's, this means that in this subsection
we will only consider
cases where the superpotential
\begin{displaymath}
W \: = \: \sum_{ij} A^{ij}(p) \phi_i \phi_j
\end{displaymath}
has $\det A \neq 0$.  (We will consider more general cases later.)

We will take the R symmetry ${\mathbb C}^{\times}_R$ to act on all 
fields with weight one (as this is the fiber version of the families
we will discuss next).  Note that this is consistent with the
constraint that $-1 \in {\mathbb C}^{\times}_R$ act trivially
on $X$, since $X$ is the ${\mathbb Z}_2$ orbifold in which the
action of $-1 \in {\mathbb C}^{\times}_R$ on the cover is quotiented.

To start with, we will assume that $n+1$ is even. From the analysis of 
\cite{cdhps}, for generic $W$ ({\it i.e.} $\det A \neq 0$),
since there is a ${\mathbb Z}_2$
gerbe structure at the critical point, and other directions are massive,
we interpret RG flow as generating two distinct points, corresponding
essentially to two different ${\mathbb Z}_2$-equivariant structures. We'll now show that we can draw the same conclusion using sheafy matrix factorizations.

\begin{itemize}

\item $n=1$

We begin with the simplest example, namely $n=1$. In this case we are working on the orbifold $[\mathbb{C}^2/\mathbb{Z}_2]$, and we choose the generic superpotential $W=xy$. Of course we have seen this example several times already, albeit without the orbifold structure.

The line $y=0$ lies within the locus $W=0$, so we have a sheafy matrix factorization ${\cal O}_{\{y=0\}}$. It's quasi-isomorphic to the traditional matrix factorization
\begin{equation}\label{lineforn=1}
 \xymatrix{
{\cal O} \ar@/_/[d]_x \\
{\cal O}(\tfrac12) \ar@/_/[u]_y
}
\end{equation}
where ${\cal O}(\tfrac12)$ denotes the
trivial line bundle with nontrivial ${\mathbb Z}_2$ equivariant structure. 
This matrix factorization is point-like - we saw this calculation in 
section~\ref{sect:pointlike}, we also do the calculation in greater generality 
in appendix~\ref{app:pointlike}. 
So this brane gives us one of our points.

To get our second point we take the sheafy matrix factorization ${\cal O}_{\{y=0\}}(\tfrac12)$.  This lives on the same line $y=0$ but carries a different $\mathbb{Z}_2$-equivariant structure. Obviously this brane is also point-like, and we claim that it represents a different point from the previous one. To justify this, we will show that
$${\rm Ext}^*_{MF}\left( {\cal O}_{\{y=0\}}, {\cal O}_{\{y=0\}}(\tfrac12) \right)=0 $$
so that homologically these two objects behave like two distinct points. 
This is not difficult to show:  if we were to forget the orbifold structure, 
then we already know that 
${\rm Ext}^*_{MF}\left( {\cal O}_{\{y=0\}}, {\cal O}_{\{y=0\}} \right)$ is 
1-dimensional (and lives in degree zero). If we twist the second brane, 
and then follow the $\mathbb{Z}_2$ charges carefully, we see that this 
single morphism becomes anti-invariant under the $\mathbb{Z}_2$ action. 
Therefore it doesn't descend to give a morphism on the orbifold. 
In other words, we have 
$${\cal E}xt_{MF}\left( {\cal O}_{\{y=0\}}, {\cal O}_{\{y=0\}}(\tfrac12) \right) = {\cal O}_0(\tfrac12) $$
(the twisted sky-scraper sheaf at the origin), but
$${\rm Ext}^*_{MF}\left( {\cal O}_{\{y=0\}}, {\cal O}_{\{y=0\}}(\tfrac12) \right)=\Gamma \left({\cal E}xt_{MF}\left( {\cal O}_{\{y=0\}}, {\cal O}_{\{y=0\}}(\tfrac12) \right)\right) = 0 $$
If we consider the line $x=0$ instead of $y=0$, then we get two more point-like objects, namely ${\cal O}_{\{x=0\}}$, and ${\cal O}_{\{x=0\}}(\tfrac12)$. However this doesn't give us additional points at the RG-flow endpoint, because we have quasi-isomorphisms
$${\cal O}_{\{y=0\}}\cong_q{\cal O}_{\{x=0\}}(\tfrac12)\;\;\;\;\;\;\;\mbox{and equivalently}\;\;\;\;\;\;\;{\cal O}_{\{y=0\}}(\tfrac12)\cong_q{\cal O}_{\{x=0\}}$$
To see this, observe that ${\cal O}_{\{x=0\}}(\tfrac12)$ is quasi-isomorphic to the traditional matrix factorization
\begin{equation}\label{lineforn=1equivalent}
 \xymatrix{
{\cal O}(\tfrac12) \ar@/_/[d]_y \\
{\cal O} \ar@/_/[u]_x
}
\end{equation}
Obviously this is closely related to the matrix factorization (\ref{lineforn=1}), indeed it appears to be a shift of it. However, despite appearances, it is in fact \textit{exactly} the same matrix factorization as (\ref{lineforn=1}). This is a slightly subtle point, and we need to recall our discussion of R symmetry from section~\ref{sect:physicsanalysis}. When we work on an ordinary manifold (rather than an orbifold), we require that the subgroup $\mathbb{Z}_2^R\subset \mathbb{C}^\times_R$ acts trivially. Consequently, every matrix factorization decomposes into a pair of eigensheaves (a brane and an anti-brane), which is why we write our matrix factorizations as pairs of sheaves. However in our current example the subgroup $\mathbb{Z}_2^R$ does not act trivally on the orbifold chart $\mathbb{C}^2$ (it only acts trivially up to gauge transformations), 
and so it is not true that matrix factorizations split up into 
$\mathbb{Z}_2^R$-eigensheaves on the covering space\footnote{If the reader 
prefers to avoid this issue, there is a solution. Instead of working on the 
chart $[\mathbb{C}^2_{x,y}/\mathbb{Z}_2]$, use the alternative chart 
$[\mathbb{C}^2_{x,y}\times \mathbb{C}^*_p \,/\, \mathbb{C}^*]$, where the 
$\mathbb{C}^*$ acts with weight 1 on $x$ and $y$ and weight $-2$ on $p$, and 
the superpotential is $W=xyp$. This is is an equally valid coordinate system 
for the same model, and has the advantage that we can choose the R-charges of 
$x$, $y$ and $p$ to be 0,0 and 2 respectively, so that $\mathbb{Z}_2^R$ acts 
honestly trivially.}.  So our notation is misleading us; we sould have 
written both (\ref{lineforn=1}) and (\ref{lineforn=1equivalent}) as

$${\cal E} = {\cal O}\oplus {\cal O}(\tfrac12) \hspace{1cm}
d_{\cal E} =\begin{pmatrix} 0 & y \\ x & 0 \end{pmatrix}$$
whence they are manifestly the same.  We continue to use the ``up-and-down'' notation because it is more customary and more compact.

The goal of this paper is to understand D-brane probes of Landau-Ginzburg
models which generically correspond to branched double covers.
This case is the prototype for generic smooth points in such double covers;
the fact that there are two matrix factorizations corresponds to the fact
we have a double cover.

\item $n=3$ 

Now set $n=3$, so we work on $[\mathbb{C}^4 / \mathbb{Z}_2]$ with the generic superpotential $W=xy + zw$. We again look for point-like sheafy matrix factorizations.

In the $n=1$ case we used the lines $x=0$ and $y=0$, the correct 
generalization of these are given by planes in $\mathbb{C}^4$ which are 
isotropic, {\it i.e.} they lie in the locus $W=0$.  
There are infinitely-many such planes, but they come in two families each 
indexed by $\mathbb{P}^1$: for each $\alpha \in \mathbb P^1$ we have the lines
$$\{ x/w = -z/y = \alpha \} \hspace{1cm}\mbox{and}\hspace{1cm} \{ x/z = -y/w = \alpha \}$$
For concreteness, let's choose the plane $U = \{x=z=0\}$. Then we have a sheafy matrix factorization ${\cal O}_U$, or if we prefer, a quasi-isomorphic traditional matrix factorization
\begin{equation}\label{lineforn=3}
\xymatrix{
{\cal O}^2 \ar@/_/[d]_{\scriptsize \left[ \begin{array}{cc}
y & -z \\ w & x \end{array} \right]} \\
{\cal O}(\tfrac12)^2 \ar@/_/[u]_{\scriptsize \left[ \begin{array}{cc}
x & z \\ -w & y \end{array} \right] }
}
\end{equation}
(As we discussed above, we should really write this as ${\cal O}^2\oplus {\cal O}(\tfrac12)^2$ with an endomorphism given by a $4\times 4$ matrix.)

One can now verify explicitly that this matrix factorization is point-like, we present the details in appendix~\ref{app:pointlike}. More specifically, it behaves like an isolated point, not embedded in a higher-dimensional space, and thus it has no non-trivial deformations: it is rigid.

Now suppose we vary $U$ within the $\mathbb{P}^1$, giving a family of sheafy matrix factorizations.  Since the matrix factorizations have no deformations, they all must be isomorphic, defining the same object of the category. So this family of isotropic planes contributes a single point at the RG-flow endpoint.

Now we proceed just as we did in the $n=1$ case. By choosing the non-trivial $\mathbb{Z}_2$-equivarant structure, we get a second point-like brane ${\cal O}_U(\tfrac12)$, and the same argument as we gave before shows that this really does represent a distinct point. If we instead choose an istropic plane $U'$ from the the other $\mathbb{P}^1$ family, we get the same two point-like objects, but just as before we have to flip the $\mathbb{Z}_2$-equariant stuctures. To see this, choose the plane $U' = \{x=w=0\}$. Then ${\cal O}_{U'}(\tfrac12)$ is quasi-isomorphic to the traditional matrix factorization
\begin{displaymath}
\xymatrix{
{\cal O}^2 \ar@/_/[d]_{\scriptsize \left[ \begin{array}{cc}
x & w \\ -z & y \end{array} \right]} \\
{\cal O}(\tfrac12)^2 \ar@/_/[u]_{\scriptsize \left[ \begin{array}{cc}
y & -w \\ z & x \end{array} \right] }
}
\end{displaymath}
This is isomorphic to (\ref{lineforn=3}) using the isomorphism where we swap the two basis vectors in both bundles.

\item Higher even cases

As long as $n+1$ is even, and the superpotential is generic, the description continues essentially unchanged. We look for subspaces $U\subset \mathbb{C}^{n+1}$ which are isotropic and have maximal dimension, which will be $(n+1)/2$ for non-degenerate $W$. For every such $U$ we have two  sheafy matrix factorizations ${\cal O}_U$ and ${\cal O}_U(\tfrac12)$, and they behave categorically like two distinct isolated points. 

Further, it is a standard fact that the maximal isotropic subspaces come in two disjoint connected families, as we saw explicitly for $n=3$ (it was trivial for $n=1$). If $U$ and $U'$ are in the same connected family then ${\cal O}_U$ and ${\cal O}_{U'}$ are quasi-isomorphic, since as a matrix factorization ${\cal O}_U$ has no deformations, and if $U$ and $U'$ live in different families then ${\cal O}_U$ is quasi-isomorphic to ${\cal O}_{U'}(\tfrac12)$. 

\item Odd cases

We'll now briefly discuss cases where $n+1$ is odd. We can produce point-like sheafy matrix factorizations in exactly the same way, by finding maximal isotropic subspaces $U\subset \mathbb{C}^{n+1}$ (for generic $W$ these will have dimension $n/2$). Then each ${\cal O}_U$ or ${\cal O}_U(\tfrac12)$  defines a sheafy matrix factorization that behaves like an isolated point.

 However, in the odd case these subspaces form a single connected family, so all the ${\cal O}_U$ are quasi-isomorphic. Furthermore, we claim that we also have a quasi-isomorphism between  ${\cal O}_U$ and  ${\cal O}_U(\tfrac12)$. So up to quasi-isomorphism we have only one point-like object, and thus after RG-flow we will see a single isolated point. This agrees with the conclusions of \cite{cdhps}.

 Let's illustate these claims in the simplest case, when $n=0$. We work on the space $[\mathbb{C}/\mathbb{Z}_2]$, with the superpotential $W=x^2$. There is only one maximal isotropic subspace $U$, namely the zero-dimensional subspace! So we do indeed see a single connected family. We have a point-like sheafy matrix ${\cal O}_0$ (the sky-scraper sheaf at the origin), and it's quasi-isomorphic to the traditional matrix factorization
\begin{displaymath}
\xymatrix{
{\cal O} \ar@/_/[d]_{x} \\
{\cal O}(\tfrac12) \ar@/_/[u]_{x}
}
\end{displaymath}
This is invariant under twisting by ${\cal O}(\tfrac12)$ (since the ordering of the pair of bundles is irrelevant), and hence ${\cal O}_0$ and ${\cal O}_0(\tfrac12)$ are quasi-isomorphic. 

As a quick aside, let's note a subtlety of the calculation of point-likeness in this case. We see immediately that
$${\cal E}xt_{MF}( {\cal O}_0, {\cal O}_0 ) = {\cal O}_0 \oplus {\cal O}_0(\tfrac12) $$
and so ${\rm Ext}^*_{MF}( {\cal O}_0, {\cal O}_0 )$ is indeed 1-dimensional as required. However, the orbifold structure is crucial here - if we do the calculation on the un-orbifolded vector space then we would get a 2-dimensional space, and the matrix factorization would not behave like an isolated point.\footnote{One might guess that it behaves like a point in a 1-dimensional space, but this is also wrong - the algebra structure on the Ext groups is incorrect.} This phenomenon occurs in all the odd cases, it's related to the fact that in the odd case it's impossible to put a $\mathbb{C}^*_R$ symmetry on the un-orbifolded model.

It is straight-forward to do a similar explicit analysis of the $n=2$ case, we leave this as an exercise.

\end{itemize}

\subsubsection{${\mathbb P}^3[2,2]$}  \label{sect:firstfamily}

Now, consider the GLSM describing ${\mathbb P}^3[2,2]$, a complete
intersection of the two quadrics $q_1$, $q_2$. At large radius, the GLSM describes a complete intersection of two quadrics,
which is an elliptic curve.
At the Landau-Ginzburg point, we have a Landau-Ginzburg model on 
\begin{displaymath}
X \: = \:
{\rm Tot}\,\left( {\cal O}(\tfrac12)^4 \: \longrightarrow \: {\mathbb P}^1_{[2,2]}
\right)
\end{displaymath}
with superpotential $W = p_1 q_1 + p_2 q_2$, where $p_1$, $p_2$
are homogeneous coordinates on ${\mathbb P}^1_{[2,2]}$. If we fix a point $p\in\mathbb{P}^1_{[2,2]}$, then on the fiber over $p$ we simply have a quadratic superpotential on an orbifold $[\mathbb{C}^4 / \mathbb{Z}_2]$. For generic points $p$ this superpotential will be non-degenerate, and we see exactly the $n=3$ case of the single quadric examples considered in the previous section. However at four points the superpotential becomes degenerate, dropping from rank 4 to rank 3. Consequently, as discussed in section~\ref{sect:glsm:nc} and \cite{cdhps},  the model should RG flow to a nonlinear sigma
model on double cover of ${\mathbb P}^1$, branched over these four points. This branched double cover is an elliptic curve; in fact it turns out to be isomorphic to the first elliptic curve.

Let us see how this IR branched double cover appears using
matrix factorizations in the Landau-Ginzburg model.

Our D-brane probes will now consist of sheafy matrix factorizations supported
in the fibers over points in the base
${\mathbb P}^1_{[2,2]}$.  Specifically, they are of the form ${\cal O}_U$ or ${\cal O}_U(\tfrac12)$, where $U$ is an isotropic subspace of a single fiber.

Let us fix a generic point $p$ in the base. As we saw in the previous section, 
the possible subspaces $U$ in the fiber over $p$ come in two 
$\mathbb{P}^1$-families. We also saw that, within a fixed fiber, 
the objects ${\cal O}_U$ and ${\cal O}_U(\tfrac12)$ are pointlike, 
and that up-to-quasi-isomorphism we only get two pointlike objects. 
In our current example the fiber sits inside a family, so we need to repeat 
the calculations taking into account the directions which are transverse to the fiber. 
But we can certainly use the results of the previous section as a guide.

Fortunately, by the general argument at the end of appendix~\ref{app:Extgroups}, the calculations are only sensitive to a first-order neighborhood of the 
fiber. So we may reduce to simple local model, namely
$$X \: = \:[\mathbb{C}^4_{x,y,z,w}\,/\, \mathbb{Z}_2]\times\mathbb{C}_p \hspace{2cm} W = xy + zw $$
Then one possible $U$ is given by the subspace $\{x=z=p=0\}$. We calculate in 
appendix~\ref{app:pointlike} that ${\cal O}_U$ is set-theoretically supported at the origin, and homologically behaves like a point living in a 1-dimensional space. Furthermore, this single degree-of-freedom corresponds to the $p$ 
coordinate, and therefore deforming $U$ within a fiber cannot change the quasi-isomorphism class, just as before.  It's also straight-forward to verify that ${\cal O}_U$ and ${\cal O}_U(\tfrac12)$ are distinct, and that if $U'$ lives in the other $\mathbb{P}^1$ family of subspaces over $p$ then  ${\cal O}_{U'}$ is quasi-isomorphic to ${\cal O}_U(\tfrac12)$. So we have moduli space of D0-branes and it forms a (trivial) double cover of the base $\mathbb{C}_p$.

We pause to compare this analysis with the one in \cite[section 3.6]{paul2}.  Aspinwall and Plesser study $\mathbb P^7[2,2,2,2]$ and the associated double cover of $\mathbb P^3$ which we will treat more fully later, but we are already in a position to clarify an issue they encounter.  They see a two D0-branes for each point of the $\mathbb P^3$, which suggests a double cover, but they see no monodromy as the point in $\mathbb P^3$ varies, which suggests a trivial (disconnected) double cover.  What is happening is that their D0-branes are traditional (vector bundle) matrix factorizations, {\it not} quasi-isomorphic to ${\cal O}_U$ and ${\cal O}_U(\tfrac12)$ where $U$ is a maximal isotropic space in the fiber, but rather to ${\cal O}_0$ and ${\cal O}_0(\tfrac12)$, where 0 is the origin in the fiber.  These in turn are both quasi-isomorphic to a direct sum of several copies of ${\cal O}_U \oplus {\cal O}_U(\tfrac12)$, so they are not seeing the two points of the double cover separately, but several copies of both together; this explains why they see no monodromy.

Returning to our analysis, let's see what happens as we approach one of the non-generic points 
where the superpotential degenerates.  Physically, in the GLSM, at these
points some minimally-charged fields become light, and the gerbe-based analysis
is no longer applicable.  Instead of a pair of points, we expect only 
a single point.  The mathematics is as follows.
A local-coordinate picture of this situation is given by
\begin{displaymath}
X \: = \: [ {\mathbb C}_{x,y,z,w}^4/{\mathbb Z}_2] \times {\mathbb C}_p \hspace{2cm} W = xy + z^2 - pw^2
\end{displaymath}
For a fixed $p$, the possible subspaces $U$ come in two families indexed by $\alpha \in \mathbb P^1$:
$$\{ x/(z-\sqrt{p}w) = -(z+\sqrt{p}w)/y = \alpha \}
\hspace{0.5cm}\mbox{and}\hspace{0.5cm}
\{ x/(z+\sqrt{p}w) = -(z-\sqrt{p}w)/y = \alpha \}$$
When $p$ goes to zero the two families coincide, so immediately it is 
reasonable to guess that the moduli space of D0-branes forms a 
\emph{branched} double cover of $\mathbb{C}_p$. However there are still 
some things to check, namely that we really do have only one point-like 
object over $p=0$, and that this object still behaves like a point in a 
smooth one-dimensional space.

To make life easier for ourselves, let's ignore the $x$ and $y$ directions. 
These are massive, and decoupled, so we can safely ignore them without affecting the results. So we can study the model
\begin{displaymath}
X \: = \: [ {\mathbb C}_{z,w}^2/{\mathbb Z}_2] \times {\mathbb C}_p \hspace{2cm} W =  z^2 - pw^2
\end{displaymath}
For general $p$, we have two isotropic subspaces (in fact lines) in the fiber, 
given by $\{z =\pm\sqrt{p}w\}$. At $p=0$, they coincide as the line $U=\{ p =z=0\}$. We need to study ${\cal O}_U$. 

A quasi-isomorphic traditional matrix factorization is given by
\begin{equation}\label{lineatp=0}
\xymatrix{
{\cal O}\oplus{\cal O}(\tfrac12) \ar@/_/[d]_{\scriptsize \left[ \begin{array}{cc}
z & -p \\ -w^2 & z \end{array} \right]} \\
{\cal O}(\tfrac12)\oplus{\cal O}  \ar@/_/[u]_{\scriptsize \left[ \begin{array}{cc}
z & p \\ w^2 & z \end{array} \right] }
}
\end{equation}
So ${\rm R}{\cal H}om({\cal O}_U, {\cal O}_U)$ is given by
$$\xymatrix{
{\cal O}_U  &
{\cal O}_U(\tfrac12) \oplus{\cal O}_U \ar[l]_(.65){[0,-w^2]}&
{\cal O}_U(\tfrac12)\ar[l]_(.35){[w^2, 0]^T}
} $$
(we have unrolled it for clarity). Taking homology we get that 
${\cal E}xt_{MF}({\cal O}_U, {\cal O}_U)$ is, as a $\mathbb{C}[z,w,p]$-module,
$$\mathbb{C}[w]/(w^2)\;\;\oplus\;\;\mathbb{C}[w]/(w^2)(\tfrac12)$$
As a sheaf, $\mathbb{C}[w]/(w^2)$ is a fat point, thickened in the $w$-direction up the fiber, and the other summand is the same sheaf with the twisted $\mathbb{Z}_2$-equivariant structure. 

Now recall that $w$ is anti-invariant under the $\mathbb{Z}_2$ action. Therefore when we take global sections the element $w$ in the first summand disappears, but the element $w$ in the second (twisted) summand survives. So we've computed
$${\rm Ext}_{MF}({\cal O}_U, {\cal O}_U)  = \mathbb{C}\oplus \mathbb{C} $$
If we keep track of the R-charges we can check that it's in fact $\mathbb C \oplus \mathbb C[-1]$.  So $\mathcal{O}_U$ still looks like a point in a smooth 1-dimensional space. Just as before, this implies that deforming $U$ within the fiber cannot change the isomorphism class of the matrix factorization, and it is also easy to check that, up to isomorphism, (\ref{lineatp=0}) is invariant under twisting by $\mathcal{O}(\tfrac12)$. So we really do only have a single pointlike object over $p=0$, and we are justified in concluding that moduli space of D0-branes is a branched double cover of $\mathbb{C}_p$. Extrapolating to the global model, we conclude that the moduli space of D0-branes is a branched double cover of $\mathbb{P}^1$, as claimed.

Notice that over the branch points the rank of $W$ becomes odd, and indeed the category of matrix factorizations looks more like it does in the odd-dimensional single-quadric examples that we analyzed in the previous section. But to calculate the Ext algebra of ${\cal O}_U$ and find that it was point-like, we couldn't just work within the single fiber over the branch point -- we had to take the transverse directions into account.

For completeness, let us also describe how to modify the RG functor
described in section~\ref{sect:RG:functor} to apply to this case. 
We will work in our simplified local model, 
$X =  [ {\mathbb C}_{z,w}^2/{\mathbb Z}_2] \times {\mathbb C}_p$, 
but the generalization to the global model is straight-forward (modulo the 
gluing issues to be discussed in section \ref{sect:gluing}).

Let ${\mathbb C}_{q}$ be the line that double-covers
${\mathbb C}_{p}$, {\it i.e.}
$q^2 = p$.  Let $Y$ be the associated double cover of $X$,
{\it i.e.}
\begin{displaymath}
Y \: = \: [{\mathbb C}_{z,w}^2 / {\mathbb Z}_2] \times {\mathbb C}_{q},
\: \: \:
W \: = \: z^2 \: - \: q^2 w^2
\end{displaymath}
Pick a sheafy matrix factorization on $(X,W)$. The functor acts as follows:
\begin{enumerate}
\item Replace it with a quasi-isomorphic traditional matrix factorization
built from vector bundles.
\item Pull up the matrix factorization to $Y$.
(This operation commutes with the first, so they can be done in either order.)
\item Restrict to the locus $\{ z = qw \}$.
\item Push down to ${\mathbb C}_q$.
\end{enumerate}

Let's apply this to one of our pointlike sheafy matrix factorizations ${\cal O}_U$.
Let $q_0$ be a fixed complex number, and let $U$ be the line $\{z=q_0w\}$ sitting in the fiber over
the point $p = q_0^2$.  We need to find a
traditional matrix factorization quasi-isomorphic to $\mathcal{O}_U$.  First, write
\begin{displaymath}
W \: = \: (z - q w)(z + q w) \: + \: (q^2 - p) w^2
\end{displaymath}
and then follow the usual recipe.  We take the Koszul resolution
of the torsion sheaf over $U=\{z = q_0 w,\, p = q_0^2 \}$,
which is given by
\begin{displaymath}
{\cal O}(\tfrac12) \: \longrightarrow \: {\cal O}(\tfrac12) \oplus {\cal O}
\: \longrightarrow \: {\cal O}
\end{displaymath}
Then we add backwards maps, to get the matrix factorization
\begin{displaymath}
\xymatrix{
{\cal O} \oplus {\cal O}(\tfrac12)
\ar@/_/[d]_{\scriptsize
\left[ \begin{array}{cc}
z - q_0 w & w^2 \\
p - q_0^2 & z + q_0 w \end{array} \right] } \\
{\cal O}(\tfrac12) \oplus {\cal O}
\ar@/_/[u]_{\scriptsize
\left[ \begin{array}{cc}
z + q_0 w & - w^2 \\
-p + q_0^2 & z - q_0 w \end{array} \right] }
}
\end{displaymath}
Pull this up to $Y$ and restrict to $\{z = q w\}$, to get
a matrix factorization of $W = 0$
\begin{displaymath}
\xymatrix{
{\cal O} \oplus {\cal O}(\tfrac12) 
\ar@/_/[d]_{\scriptsize
\left[ \begin{array}{cc}
(q - q_0) w & w^2 \\
q_0^2 - q^2 & (q + q_0) w \end{array} \right] } \\
{\cal O}(\tfrac12) \oplus {\cal O}
\ar@/_/[u]_{\scriptsize
\left[ \begin{array}{cc}
(q + q_0) w & -w^2 \\
q_0^2 - q^2 & (q - q_0) w \end{array} \right] }
}
\end{displaymath}
The homology of this consists of 
\begin{displaymath}
{\mathbb C}[q,w]/(w, q - q_0), \: \: \:
{\mathbb C}[q,w](\tfrac12) / (w, q + q_0)
\end{displaymath}
After pushing down to ${\mathbb C}_{q}$ (taking ${\mathbb Z}_2$
invariants), we get a skyscraper sheaf at $q = q_0$,
exactly as desired.

\subsubsection{${\mathbb P}^{2g+1}[2,2]$}

An example discussed in section~4.1 of \cite{cdhps} is a complete
intersection of two quadrics in ${\mathbb P}^{2g+1}$.
When $g \neq 1$, this is not Calabi-Yau, but we can still consider
B-type D-branes and matrix factorizations in the corresponding
untwisted physical theory.

As discussed in \cite{cdhps}, at large radius the GLSM describes the
complete intersection above, and at the Landau-Ginzburg point, describes
a Landau-Ginzburg model on
\begin{displaymath}
{\rm Tot}\left( {\cal O}(\tfrac12)^{2g+2} \: \longrightarrow \: 
{\mathbb P}^1_{[2,2]} \right)
\end{displaymath}
which RG flows to a branched double cover of ${\mathbb P}^1$, giving a 
hyperelliptic curve of genus $g$.

The analysis of matrix factorizations in this example is identical
to that of matrix factorizations in the theory associated to
${\mathbb P}^3[2,2]$, discussed in the previous section. It concludes that we can see this hyperelliptic curves as a moduli space of pointlike sheafy matrix factorizations.

\subsubsection{${\mathbb P}^5[2,2,2]$}  \label{sect:secondfamily}

A closely related example involves the GLSM associated to
${\mathbb P}^5[2,2,2]$, discussed in section~4.3 of \cite{cdhps}.
Here, at the Landau-Ginzburg point, one has a Landau-Ginzburg model on
the total space of
\begin{displaymath}
{\rm Tot}\,\left( {\cal O}(\tfrac12)^6 \: \longrightarrow \: {\mathbb P}^2_{[2,2,2]}
\right)
\end{displaymath}
It was argued in \cite{cdhps} that this example RG flows to a branched
double cover of ${\mathbb P}^2$, branched over a degree 6 locus.
We can analyze matrix factorizations in this model in the same fashion
as above -- at generic points on ${\mathbb P}^2$, we see two pointlike matrix factorizations, but over the branch locus, there is only one. Furthermore, these matrix factorizations behave like points in a smooth 2-dimensional space, even over the branch locus. Hence the moduli space of D0-branes really is the branched double cover.

\subsubsection{Degree 4 del Pezzo ${\mathbb P}^4[2,2]$}
\label{sect:deg-4-dP}

The GLSM for this example was discussed in section~4.4 of
\cite{cdhps}, and is somewhat different from the previous examples.
Here, we have a Landau-Ginzburg model on the total space of
\begin{displaymath}
{\rm Tot}\,\left( {\cal O}(\tfrac12)^5 \: \longrightarrow \:
{\mathbb P}^1_{[2,2]} \right)
\end{displaymath}

It was argued in \cite{cdhps} that this should RG flow to a
single copy of ${\mathbb P}^1$, with five ${\mathbb Z}_2$ orbifold points. These orbifold points occur at the points where the fiber-wise superpotential becomes degenerate.

We can construct matrix factorizations as before, but here the prototypical
fiber example from section~\ref{sect:onequadric} is the case 
$V = {\mathbb C}^3$,
$W = xy + z^2$.  In this case, there is a single point-like matrix factorization, and if we analyze the transverse directions as well (the calculation is very similar to the one done in appendix~\ref{app:pointlike}) we see a point moving in a smooth one-dimensional space. So near generic points, the moduli space of D0-branes is a single-cover of $\mathbb{P}^1$, in agreement with  \cite{cdhps}.

Unfortunately at degenerate points this kind of analysis is not much use, for the following reason: the moduli space of D0-branes is never an orbifold! This is a familiar fact, D0-brane probes of orbifolds do not recover the orbifold, 
they recover a manifold that resolves the orbifold singularity. 
The mathematical reason for it is simple, it's because sheaves 
(or complexes, or matrix factorizations) never have finite non-trivial 
automorphism groups.

It is however possible to show that the \emph{category} of sheafy matrix factorizations is equivalent to the derived category of coherent sheaves on this $\mathbb{P}^1$-with-orbifold-points, and to construct an RG-flow functor between the two. But we shall not pursue this here.

Very similar remarks apply to the example ${\mathbb P}^6[2,2,2]$ discussed
in section~4.5 of \cite{cdhps}.

\subsection{Global gluing issues}
\label{sect:gluing}

So far in our analyses of D-brane probes of Landau-Ginzburg models
corresponding to various complete intersections of quadrics, we have
worked locally over the base space, in a Born-Oppenheimer approximation.

However, there is also some interesting global information that
can be extracted, pertinent to the $B$ field.  In this section, we will
outline such global gluing issues.

In general, in our D0-brane moduli space, over each point of the
base space we have described a D0-brane by picking an isotropic
subspace $U$.  The set of isotropic subspaces of a vector space with a 
quadratic form is known as an isotropic Grassmannian\footnote{
As isotropic Grassmannians are not commonly used in the physics
community, we collect here a handful of pertinent facts.
The isotropic Grassmannian denoted $OGr(p,n)$ is defined to be the
space of $p$-dimensional complex vector subspaces $U$ of a fixed
$n$-dimensional complex 
vector space $V$, such that $U$ is isotropic with respect
to a fixed symmetric bilinear form.
If $n$ is even, then $OGr(n/2,n)$ always has two components.
For example, the components of $OGr(2,4)$ are copies of ${\mathbb P}^1$;
the components of $OGr(3,6)$ are copies of ${\mathbb P}^3$;
the components of $OGr(4,8)$ are copies of a six-dimensional quadric.
In odd-dimensional cases, $OGr(n,2n+1)$ is isomorphic to one component
of $OGr(n+1,2n+2)$.
}, so the set of
choices of $U$'s over every point forms a bundle over the base  
whose fibers are copies of an isotropic Grassmannian.

In general, that bundle of isotropic Grassmannians will not have a global
section.  Since the D0-branes given by different choices of $U$ are isomorphic
but not canonically so, we can only glue together our choices of $U$'s up 
to an overall (${\mathbb C}^{\times}$) automorphism.  Physically, this
obstruction to gluing D-branes globally corresponds precisely to having
a nontrivial $B$ field.  Recall that in the presence of a topologically
nontrivial $B$ field, the Chan-Paton factors couple to a twisted bundle,
where the twisting is determined by the topological class of the $B$
field.  Here, the ${\mathbb C}^{\times}$ automorphism obstructing
a global gluing precisely corresponds to a choice of
${\mathbb C}^{\times}$ gerbe, and hence a topologically nontrivial
$B$ field.

That said, in the case of the Landau-Ginzburg model
corresponding to ${\mathbb P}^3[2,2]$, there will be a section.
Briefly, there are two quadratic forms $q_0$, $q_1$ on ${\mathbb P}^3$,
and $X = \{q_0 = q_1 = 0\}$ is an elliptic curve.  For any
point $x \in X$, for any quadratic form
\begin{displaymath}
q \: = \: a_0 q_0 \: + \: a_1 q_1
\end{displaymath}
there is at least one line (singular quadrics) and typically two (smooth
quadrics) on the quadratic surface containing $x$, hence the bundle
of isotropic Grassmannians has a section in this case, so the
$B$ field is topologically trivial.

More generally, for ${\mathbb P}^{2g+1}[2,2]$ there will also be a section
of the bundle of isotropic Grassmannians, and hence a topologically-trivial
$B$ field, which can be argued by choosing a ${\mathbb P}^{g-1}$ on the
intersection of two quadrics.  However, this argument only works for the
intersection of two quadrics; for intersections of three or more quadrics,
in general the $B$ field will be topologically nontrivial.

\section{D-brane probes of nc spaces}
\label{sect:nc}

So far we have discussed D-brane probes of Landau-Ginzburg models that
are believed \cite{cdhps} to flow to nonlinear sigma models on branched
double covers, and we have recovered precisely that structure in the
D-brane probes.

Next, we consider D-brane probes of Landau-Ginzburg models that are
believed \cite{cdhps} to flow to noncommutative resolutions of
branched double covers.  At the level of Landau-Ginzburg models,
these examples are very similar to the branched double cover examples
discussed previously.  The primary difference is that the previous
analysis yields a singular branched double cover, whereas 
other analyses of the Landau-Ginzburg model (or UV GLSM) suggest the
model should be nonsingular.  

We will repeat exactly the same analysis as for D-brane probes of
branched double covers.  In the case of noncommutative resolutions,
the D-brane probes will yield (non-K\"ahler) small resolutions of the
singularities.  As discussed elsewhere, such non-K\"ahler resolutions
cannot\footnote{
In particular, these are not believed to be generalized 
Calabi-Yau in the sense of Hitchin's generalized complex geometry,
so their non-K\"ahler property is inconsistent with the structure of
a (2,2) SCFT.
} themselves consistently be the target of (2,2) supersymmetric
nonlinear sigma models, and so we do not interpret this result to mean
that the closed string theory has target space a non-K\"ahler small
resolution.  Instead, we recall that
D-brane probe
moduli spaces are always, by construction, spaces, even when the
closed string CFT does not admit a geometric interpretation
(see {\it e.g.} \cite{dgm,dks} for discussion of other examples).

\subsection{${\mathbb P}^7[2,2,2,2]$}

Now, let us turn to the nc spaces arising in \cite{cdhps} as homological
projective duals to ${\mathbb P}^7[2,2,2,2]$.  As discussed
in \cite{cdhps} and section~\ref{sect:glsm:nc}, we have a 
Landau-Ginzburg model on
\begin{displaymath}
X \: = \: {\rm Tot}\left( {\cal O}(-\tfrac12)^8 \: \longrightarrow \:
{\mathbb P}^3_{[2,2,2,2]} \right)
\end{displaymath}
(a fiber bundle over ${\mathbb P}^3$ with fibers
$[ {\mathbb C}^8 / {\mathbb Z}_2 ]$), and superpotential
\begin{displaymath}
W \: = \: \sum_a p_a G_a(\phi) \: = \:
\sum_{ij} \phi_i \phi_j A^{ij}(p)
\end{displaymath}
The noncommutative resolution is described by
$({\mathbb P}^3, {\cal B})$, where ${\cal B} \in {\rm Coh}({\mathbb P}^3)$
is the sheaf of even parts of Clifford algebras associated with the
universal quadric defined by $W$, and describes the possible
matrix factorizations in this case.

As in the last section, we will work locally, and furthermore,
we will work with a slightly simplified toy model of the
noncommutative 
resolution $({\mathbb P}^3,{\cal B})$, in order to understand D-brane probes.
Specifically, consider a Landau-Ginzburg model on
\begin{displaymath}
{\rm Tot}\left( {\cal O} \oplus {\cal O} \: \longrightarrow \:
{\mathbb C}^3 \right)/{\mathbb Z}_2
\end{displaymath}
where the ${\mathbb Z}_2$ acts trivially on the base, and by sign flips along
the fibers,
with superpotential
\begin{displaymath}
W \: = \: \sum_{i,j=1}^2 A^{ij}(a,b,c) \phi_i \phi_j
\end{displaymath}
where 
\begin{displaymath}
(A^{ij}) \: = \: \left[ \begin{array}{cc}
a & b/2 \\ b/2 & c \end{array} \right]
\end{displaymath}
so that 
\begin{displaymath}
W \: = \: a \phi_1^2 \: + \: b \phi_1 \phi_2 \: + \: c \phi_2^2
\end{displaymath}
In this toy model, we have a double cover of 
\begin{displaymath}
{\mathbb C}^3 \: = \: {\rm Spec}\, {\mathbb C}[a,b,c]
\end{displaymath}
branched over the locus $\det A \propto b^2 - 4ac$.
(That branched double cover could be described algebraically as
the conifold $b^2 - 4ac + d^2 = 0$ in ${\mathbb C}^4$.)

Now, let us consider matrix factorizations in this toy model,
defined by sheaves over points in ${\mathbb C}^3$.
Define
\begin{displaymath}
\phi_{\pm} \: \equiv \:
2a \phi_1 \: + \: b \phi_2 \: \pm \: \sqrt{b^2 - 4ac} \, \phi_2
\end{displaymath}
and note that $\phi_+ \phi_- \propto W$.
(We can absorb the proportionality factor into an automorphism of
one of the ${\cal O}_p$'s, and so we omit it.)
If $F$ is the fiber over the point with coordinates $(a,b,c)$
then for generic points on ${\mathbb C}^3$ we have two matrix factorizations:
\begin{displaymath}
\xymatrix{
{\cal O}_F \ar@/_/[d]_{\phi_+} \\
{\cal O}_F(\tfrac12) \ar@/_/[u]_{\phi_-}
}, \: \: \:
\xymatrix{
{\cal O}_F \ar@/_/[d]_{\phi_-} \\
{\cal O}_F(\tfrac12) \ar@/_/[u]_{\phi_+}
}
\end{displaymath}
which correspond to the two points on the double cover.

Along the curve $\det A = 0$, $\phi_+ = \phi_-$, so these two matrix
factorizations become a single matrix factorization, reflecting the fact
that the two sheets coincide along that curve.

At the point $a=b=c=0$, something more interesting happens.
At that point, then for any linear combination $\phi$ of $\phi_1$, $\phi_2$,
we have two matrix factorizations:
\begin{displaymath}
\xymatrix{
{\cal O}_F \ar@/_/[d]_{0} \\
{\cal O}_F(\tfrac12) \ar@/_/[u]_{\phi}
}, \: \: \:
\xymatrix{
{\cal O}_F \ar@/_/[d]_{\phi} \\
{\cal O}_F(\tfrac12) \ar@/_/[u]_0
}
\end{displaymath}
The possible linear combinations $\phi$ are described, 
up to the rescaling afforded by automorphisms,
by a point on ${\mathbb P}^1$, so we see that the two choices above correspond
to two choices of points on ${\mathbb P}^1$, and precisely describe
the two possible small resolutions.

In an actual D-brane probe moduli space, to pick one particular
small resolution, one would need to impose a notion of stability,
as indeed happens with K\"ahler parameters in D-brane probes
of orbifolds as discussed in \cite{dgm}.  We have not picked a notion
of stability, so we should expect to see both, as indeed we have.

In the previous toy model, we saw a small resolution, but its structure
seemed to be interrelated to having only two $\phi$'s.
Let us now generalize slightly to four $\phi$'s, to illustrate how the
toy example above will generalize, and how we will again get
small resolutions.  In this case, we have 
the superpotential
\begin{displaymath}
W \: = \: \sum_{ij=1}^4 \phi_i \phi_j A^{ij}(p)
\end{displaymath}
For example, over a point such that $W$ is the smooth quadric
\begin{displaymath}
W \: = \:
\phi_1 \phi_2 \: + \: \phi_3 \phi_4
\end{displaymath}
we have the matrix factorizations
\begin{displaymath}
\xymatrix{
{\cal O}_F^2 \ar@/_/[d]_{\phi_+} \\
{\cal O}_F(\tfrac12)^2 \ar@/_/[u]_{\phi_-} 
}, \: \: \:
\xymatrix{
{\cal O}_F^2 \ar@/_/[d]_{\phi_-} \\
{\cal O}_F(\tfrac12)^2 \ar@/_/[u]_{\phi_+}
}
\end{displaymath}
where
\begin{displaymath}
\phi_{+} \: = \: \left[ \begin{array}{cc}
\phi_1 & - \phi_3 \\ \phi_4 & \phi_2 \end{array} \right], \: \: \:
\phi_- \: = \: \left[ \begin{array}{cc}
\phi_2 & \phi_3 \\ - \phi_4 & \phi_1 \end{array} \right],
\end{displaymath}
just as in the corresponding one-quadric example discussed in
section~\ref{sect:onequadric}.

Over a  point such that $W$ is the singular quadric
\begin{displaymath}
W \: = \: \phi_1 \phi_2 \: + \: \phi_3^2
\end{displaymath}
we have the matrix factorization
\begin{displaymath}
\xymatrix{
{\cal O}_F^2 \ar@/_/[d]_P \\
{\cal O}_F(\tfrac12)^2 \ar@/_/[u]_Q
}
\end{displaymath}
where
\begin{displaymath}
P \: = \: \left[ \begin{array}{cc}
\phi_1 & - \phi_3 \\ \phi_3 & \phi_2 \end{array} \right], \: \: \:
Q \: = \: \left[ \begin{array}{cc}
\phi_2 & \phi_3 \\ - \phi_3 & \phi_1 \end{array} \right],
\end{displaymath}
exactly as in the corresponding one-quadric example in
section~\ref{sect:onequadric}.  As in that section,
the matrices $P$, $Q$ are conjugate:
\begin{displaymath}
U P U^{-1} \: = \: Q \mbox{ for }
U \: = \: \left[ \begin{array}{cc}
0 & 1 \\ 1 & 0 \end{array} \right]
\end{displaymath}
so one does
not get a new matrix factorization from exchanging them.

Over a point such that $W$ is the even more singular quadric
\begin{displaymath}
W \: = \: \phi_1 \phi_2
\end{displaymath}
we have the matrix factorization
\begin{displaymath}
\xymatrix{
{\cal O}_F^2 \ar@/_/[d]_P \\
{\cal O}_F(\tfrac12)^2 \ar@/_/[u]_Q
}
\end{displaymath}
where
\begin{displaymath}
P \: = \: \left[ \begin{array}{cc}
\phi_1 & f \\ 0 & \phi_2 \end{array} \right], \: \: \:
Q \: = \: \left[ \begin{array}{cc}
\phi_2 & -f \\ 0 & \phi_1 \end{array} \right] \: = \: \phi_1 \phi_2 P^{-1},
\end{displaymath}
as in the corresponding example in section~\ref{sect:onequadric},
where $f$ is a linear function of the $\phi$'s.

Without loss of generality we can
take $f$ to depend only upon $\phi_3$, $\phi_4$.  This is because
\begin{displaymath}
\left[ \begin{array}{cc}
1 & -b \\ 0 & 1 \end{array} \right] \,
\left[ \begin{array}{cc}
\phi_1 & a \phi_1 + b \phi_2 + c \phi_3 + d \phi_4 \\
0 & \phi_2 \end{array} \right]\, 
\left[ \begin{array}{cc}
1 & -a \\ 0 & 1 \end{array} \right]
\: = \:
\left[ \begin{array}{cc}
\phi_1 & c \phi_3 + d \phi_4 \\ 0 & \phi_2 \end{array} \right],
\end{displaymath}
(and similarly for $Q$,)
hence any $\phi_1$, $\phi_2$ dependence can be reabsorbed into
automorphisms of the ${\cal O}_F$'s.  We can similarly rescale,
thus, the coefficients $c$, $d$ of $\phi_3$, $\phi_4$ act as 
homogeneous coordinates on a ${\mathbb P}^1$, which is precisely a
small resolution of this singularity.

If we swap $P$, $Q$, or equivalently\footnote{
To see that swapping and transposing are equivalent, 
conjugate by the isomorphism defined by
\begin{displaymath}
\left( \begin{array}{cc}
0 & 1 \\
1 & 0 \end{array} \right).
\end{displaymath}
} take the transpose,
then in the same fashion we recover a second small
resolution.  In order to form a moduli space in which only one of the
small resolutions are present, as discussed earlier, we must pick a stability
condition, as in \cite{dgm}.  We have not done so, hence we see all possible
small resolutions with these methods.

Then, for $2n$ $\phi$'s, we have matrix factorizations of 
$2^{n-1} \times 2^{n-1}$ matrices.

\subsection{${\mathbb P}^6[2,2,2,2]$}

Now, let us consider the GLSM associated to ${\mathbb P}^6[2,2,2,2]$,
as described in \cite[section 4.6]{cdhps}.

Here, the theory at the Landau-Ginzburg point has odd-dimensional fibers,
so instead of a branched double cover, from the analysis of
\cite{cdhps} one gets in the IR a 
single cover of ${\mathbb P}^3$ with a degree 7 hypersurface of 
${\mathbb Z}_2$ orbifolds.  In addition, this space has several ordinary
double points, and hence there is a noncommutative resolution structure.

As in our analysis of ${\mathbb P}^4[2,2]$ in section~\ref{sect:deg-4-dP},
a purely local analysis away from the locus $\{ \det A = 0 \}$ gives a 
single cover, consistent with the story above.  As the details are
nearly identical to what has been described elsewhere, we omit them.
However, the structure at the locus $\{ \det A = 0 \}$ is more complicated,
and cannot be properly straightened out without imposing some sort of
stability condition.

Therefore, our results in this case are only partial.  We do see a structure
compatible with that claimed in \cite{cdhps}, but we also cannot completely
independently verify all aspects of the description given there.

\subsection{Summary}

So far we have seen that D-brane probes of the nc resolutions
appearing in \cite{cdhps} yield (non-K\"ahler) small resolutions
of singular spaces.  In particular, the D-brane probes see a different
space than the closed string sector.  This is neither novel nor
unanticipated:
\begin{itemize}
\item By construction, D-brane probe spaces are spaces, even when the
closed string sector does not have a geometric interpretation.
\item Another example appears in \cite[section 6.2]{kps}, 
\cite[section 8]{dks}, where D-brane probes of orbifolds were discussed.
It is an old fact that D-brane probes of orbifolds see (not necessarily
Calabi-Yau) resolutions of the singularity.  However, the closed string
sector is a sigma model on a simple quotient stack, not a 
resolution of a quotient space.
(The fact that the 
resolutions
are not necessarily Calabi-Yau is already one indication of a difference;
the fact that the structure of twisted sectors does not appear anywhere
in resolutions is another indicator.)  
\end{itemize}

\subsection{More general statements}

There are suggestions, due to B. Toen and M. Vaquie \cite{tonypriv}, 
that this structure generalizes, in the sense that D-brane probes of
noncommutative resolutions will always see (not necessarily K\"ahler)
small resolutions of the underlying singular spaces.

We will not attempt to work out the details here, but instead merely
outline their statements and a conjecture.
Toen and Vaquie (in unpublished work) define a notion of point objects,
and consider a moduli stack of deformations of those point objects.
When that moduli stack is nonempty (apparently not all nc spaces admit
point objects of their form),
it is a ${\mathbb C}^{\times}$ gerbe over a proper algebraic
space, which in the present context would translate to the statement that
it is a (not necessarily K\"ahler) small resolution.

We conjecture that when applied to noncommutative resolutions,
when that space exists, it is the D0-brane moduli space.
(Checking such a statement would require a more functorial definition of
D0-brane moduli spaces.)

We leave such considerations for future work.

\section{Conclusions}

In this paper we have used D-brane probes to study the results of
\cite{cdhps}, involving Landau-Ginzburg points
of some GLSMs that flow in the IR to nonlinear sigma models on
branched double covers (realized nonperturbatively, instead of as the
critical locus of a superpotential), and noncommutative resolutions thereof.

For cases corresponding to smooth branched double covers, our D-brane probe
moduli spaces see the branched double cover structure explicitly,
verifying the results of \cite{cdhps}.

For nongeometric cases, corresponding to noncommutative resolutions of
singular branched double covers, the D-brane probe moduli spaces are
(usually non-K\"ahler) small resolutions of the singularities.
This is not interpreted as the target space of a closed string
(2,2) SCFT, but rather illustrates how D-brane probe moduli spaces
can differ from the closed string interpretation, as also happens in
{\it e.g.} orbifolds.

The noncommutative resolutions appearing in \cite{cdhps} represent
new (2,2) SCFT's, and the investigations in this paper and in 
\cite{cdhps} represent just the beginning of work that should be
done to understand their properties.  An example of a future direction
would be to compute Gromov-Witten invariants of
noncommutative spaces, perhaps by working in a UV Landau-Ginzburg
description via A-twisted Landau-Ginzburg models 
\cite{alg22,fjr1,fjr2}.

\section{Acknowledgements}

We thank D.~Deliu,
D.-E.~Diaconescu and T.~Pantev for useful conversations.
NMA was partially supported by NSF grants DMS-0556042 and DMS-0838210 and EPSRC grant EP/G06170X/1.
EPS was supported by an Imperial College Junior Research Fellowship.
ERS was partially supported by NSF grants DMS-0705381, PHY-0755614,
and PHY-1068725.

\appendix

\section{Technical definitions}  \label{sect:tech}

As this paper is written for a physics audience, it may be useful
to review some technical definitions and lemmas involving matrix
factorizations.  We do not claim this material is new; rather, we are
merely collecting it here to make this paper self-contained.

\subsection{Homotopies}

Given two matrix factorizations
\begin{displaymath}
\xymatrix{
{\cal E}_0 \ar@/_/[d]_{P_{\cal E}} \\
{\cal E}_1 \ar@/_/[u]_{Q_{\cal E}}
}, \: \: \:
\xymatrix{
{\cal F}_0 \ar@/_/[d]_{P_{\cal F}} \\
{\cal F}_1 \ar@/_/[u]_{Q_{\cal F}}
}
\end{displaymath}
(in which $P \circ Q = W \, {\rm Id}$ and $Q \circ P = W \, {\rm Id}$),
a map between them $f: {\cal E} \rightarrow {\cal F}$
is a pair of maps $f_0: {\cal E}_0 \rightarrow
{\cal F}_0$, $f_1: {\cal E}_1 \rightarrow {\cal F}_1$ such that
$P_{\cal F} \circ f_0 = f_1 \circ P_{\cal E}$ and $Q_{\cal F} \circ f_1
= f_0 \circ Q_{\cal E}$, making the following diagram commute: 
\begin{displaymath}
\xymatrix{
{\cal E}_0 \ar[r]^{f_0} \ar@/_/[d]_{P_{\cal E}} &
{\cal F}_0 \ar@/_/[d]_{P_{\cal F}} \\ 
{\cal E}_1 \ar[r]^{f_1}  \ar@/_/[u]_{Q_{\cal E}} &
{\cal F}_1 \ar@/_/[u]_{Q_{\cal F}}
}
\end{displaymath}
Equivalently, if there is an R symmetry, if we express the two
matrix factorizations as pairs $({\cal E}, d_{\cal E})$ and
$({\cal F}, d_{\cal F})$, then a map between matrix factorizations
is an R-symmetry-invariant map $f: {\cal E} \rightarrow {\cal F}$
that commutes with the $d$'s:
\begin{displaymath}
\xymatrix{
{\cal E} \ar[r]^f \ar[d]_{d_{\cal E}} & {\cal F} \ar[d]^{d_{\cal F}} \\
{\cal E} \ar[r]^f & {\cal F}
}
\end{displaymath}

We say two maps $f, g: {\cal E} \rightarrow {\cal F}$ are homotopic
(and write $f \sim g$) if there exist maps $s: {\cal E}_0 \rightarrow
{\cal F}_1$, $t: {\cal E}_1 \rightarrow {\cal F}_0$
such that
\begin{equation}  \label{homotopyformula}
f_0 \: - \:  g_0 \: = \: Q_{\cal F} \circ s \: + \: t \circ P_{\cal E},
\: \: \:
f_1 \: - \: g_1 \: = \: s \circ Q_{\cal E} \: + \: P_{\cal F} \circ t
\end{equation}
(Just as in \cite{snowbird,al}, homotopy in the mathematical sense
above corresponds physically to BRST-exactness of the difference.)
If a map $f \sim 0$, we say $f$ is null-homotopic.
We say two matrix factorizations ${\cal E}$, ${\cal F}$ are 
homotopy-equivalent
to one another if there exist maps $F: {\cal E} \rightarrow {\cal F}$,
$G: {\cal F} \rightarrow {\cal E}$ such that
\begin{displaymath}
G \circ F \: \sim \: 1_{\cal E}, \: \: \:
F \circ G \: \sim \: 1_{\cal F}
\end{displaymath}
The maps $F$ and $G$, in this case, are known as homotopy equivalences.
As a result,
${\cal E}$ is homotopy-equivalent to 0 (the zero matrix factorization)
precisely when $1_{\cal E} \sim 0$; in such a case, we say that ${\cal E}$
is contractible. 

\subsection{Cones}

We define the cone over $f$, $C(f)$, to be the matrix factorization
\begin{displaymath}
\xymatrix{
{\cal F}_0 \oplus {\cal E}_1 \ar@/_/[d]_{\hat{P}} \\
{\cal F}_1 \oplus {\cal E}_0 \ar@/_/[u]_{\hat{Q}}
}
\end{displaymath}
where
\begin{displaymath}
\hat{P} \: = \: \left[ \begin{array}{cc}
P_{\cal F} & f_1 \\
0 & -Q_{\cal E} \end{array} \right], \: \: \:
\hat{Q} \: = \: \left[ \begin{array}{cc}
Q_{\cal F} & f_0 \\
0 & -P_{\cal E} \end{array} \right]
\end{displaymath}
It is straightforward to check that $\hat{P}\hat{Q}$ and $\hat{Q}\hat{P}$ are
both $W \, {\rm Id}$, as expected.

An iterated cone over a complex can be defined similarly.
Suppose we have a complex of matrix factorizations
\begin{displaymath}
{\cal E} \: \stackrel{f}{\longrightarrow} \: {\cal F} \: 
\stackrel{g}{\longrightarrow} \: {\cal G}
\end{displaymath}
or in detail,
\begin{displaymath}
\xymatrix{
{\cal E}_0 \ar[r]^{f_0} \ar@/_/[d]_{P_{\cal E}} &
{\cal F}_0 \ar[r]^{g_0} \ar@/_/[d]_{P_{\cal F}} &
{\cal G}_0 \ar@/_/[d]_{P_{\cal G}} \\
{\cal E}_1 \ar[r]^{f_1} \ar@/_/[u]_{Q_{\cal E}} &
{\cal F}_1 \ar[r]^{g_1} \ar@/_/[u]_{Q_{\cal F}} &
{\cal G}_1 \ar@/_/[u]_{Q_{\cal G}}
}
\end{displaymath}
The cone over the complex, $C(f,g)$ is defined to be the
matrix factorization
\begin{displaymath}
\xymatrix{
{\cal G}_0 \oplus {\cal F}_1 \oplus {\cal E}_0
\ar@/_/[d]_{\hat{P}} \\
{\cal G}_1 \oplus {\cal F}_0 \oplus {\cal E}_1
\ar@/_/[u]_{\hat{Q}}
}
\end{displaymath}
where
\begin{displaymath}
\hat{P} \: = \: \left[ \begin{array}{ccc}
P_{\cal G} & g_1 & 0 \\
0 & -Q_{\cal F} & -f_0 \\
0 & 0 & + P_{\cal E} \end{array} \right], \: \: \:
\hat{Q} \: = \: \left[ \begin{array}{ccc}
Q_{\cal G} & g_0 & 0 \\
0 & -P_{\cal F} & -f_1 \\
0 & 0 & Q_{\cal E} \end{array} \right]
\end{displaymath}
It is straightforward to check that $\hat{P}\hat{Q}$ and $\hat{Q}\hat{P}$ are
both $W \, {\rm Id}$ so long as $g_i f_i = 0$.

As the name suggests, we can also think of the iterated cone as
a cone over a cone.  Given the complex of matrix factorizations above,
the map $g$ defines a map $\tilde{g}: C(f) \rightarrow {\cal G}$,
given by
\begin{displaymath}
\xymatrix{
{\cal F}_0 \oplus {\cal E}_1 \ar@/_/[d]_{\hat{P}} \ar[r]^{[g_0,0]}
&
{\cal G}_0 \ar@/_/[d]_{P_{\cal G}} \\
{\cal F}_1 \oplus {\cal E}_0 \ar@/_/[u]_{\hat{Q}} \ar[r]^{[g_1,0]} 
&
{\cal G}_1 \ar@/_/[u]_{Q_{\cal G}}
}
\end{displaymath}
It is straightforward to check that $C(\tilde{g}) = C(f,g)$.

In the other direction, the map $f$ defines $\tilde{f}: {\cal E}[1]
\rightarrow C(g)$, as
\begin{displaymath}
\xymatrix{
{\cal E}_1 \ar@/_/[d]_{Q_{\cal E}} \ar[r]^{[0, f_1]^T }
&
{\cal G}_0 \oplus {\cal F}_1 \ar@/_/[d] \\
{\cal E}_0 \ar@/_/[u]_{P_{\cal E}} \ar[r]^{[0, -f_0]^T}
& {\cal G}_1 \oplus {\cal F}_0 \ar@/_/[u]
}
\end{displaymath}
Although $C(\tilde{f})$ is not quite identical to $C(f,g)$,
it is isomorphic, with isomorphism $F: C(\tilde{f}) \rightarrow C(f,g)$
defined by $F_0 = {\rm Id}$, $F_1 = {\rm diag}(1,1,-1)$.

In section~\ref{sect:quasiiso}, we discuss a notion of quasi-isomorphisms
between matrix factorizations.  Briefly, a map $f$ of matrix factorizations
is defined to be a quasi-isomorphism if the cone $C(f)$ is 
homotopy-equivalent to
the iterated cone over an exact sequence.

As a simple example, if $f: {\cal E} \rightarrow {\cal F}$ defines a 
homotopy equivalence, then we shall show momentarily that
$C(f)$ is homotopy-equivalent to $C(1_{\cal E})$,
hence homotopy equivalences are also quasi-isomorphisms (since their cones
are homotopy-equivalent to cones over exact sequences).  
Now, let us quickly review
how $C(f) \sim C(1_{\cal E})$.  Let $g: {\cal F} \rightarrow {\cal E}$
be the homotopy inverse, so that $f g \sim 1_{\cal F}$, $g f \sim 1_{\cal E}$
(with corresponding homotopies $s_{\cal F}$, $t_{\cal F}$, $s_{\cal E}$,
$t_{\cal E}$, respectively).
The maps $f, g$ induce $\tilde{f}: C(1_{\cal E}) \rightarrow C(f)$, 
$\tilde{g}: C(f) \rightarrow C(1_{\cal E})$, defined
as, respectively,
\begin{displaymath}
\xymatrix{
{\cal E}_0 \oplus {\cal E}_1 \ar[r]^{\scriptsize \left[ \begin{array}{cc}
f_0 & 0 \\
0 & 1 \end{array} \right] } \ar@/_/[d] 
&
{\cal F}_0 \oplus {\cal E}_1 \ar@/_/[d] \\
{\cal E}_1 \oplus {\cal E}_0 \ar@/_/[u]
\ar[r]_{\scriptsize \left[ \begin{array}{cc}
f_1 & 0 \\
0 & 1 \end{array} \right] }
&
{\cal F}_1 \oplus {\cal E}_0 \ar@/_/[u]
}, \: \: \:
\xymatrix{
{\cal F}_0 \oplus {\cal E}_1 \ar@/_/[d]
\ar[r]^{\scriptsize \left[ \begin{array}{cc}
g_0 & t_{\cal E} \\
0 & 1 \end{array} \right] }
& {\cal E}_0 \oplus {\cal E}_1 \ar@/_/[d] 
\\
{\cal F}_1 \oplus {\cal E}_0 \ar@/_/[u]
\ar[r]_{\scriptsize \left[ \begin{array}{cc}
g_1 & s_{\cal E} \\
0 & 1 \end{array} \right] }
&
{\cal E}_1 \oplus {\cal E}_0 \ar@/_/[u]
}
\end{displaymath}
It can be shown that 
$\tilde{g} \circ \tilde{f}$ is homotopic to $1_{C(1_{\cal E})}$,
and $\tilde{f} \circ \tilde{g}$ is homotopic to $1_{C(f)}$.
Putting this together, we have that $C(f) \sim C(1_{\cal E})$,
hence homotopy equivalences are also quasi-isomorphisms.

Next, we derive some small lemmas that will be useful later. 
Firstly, suppose $f: {\cal E} \to {\cal F} $ is any map, and that 
${\cal E}$ is contractible. 
Then we claim that $C(f)$ is homotopy-equivalent to ${\cal F}$.

Let $s: {\cal E}_0 \to {\cal E}_1$ and  $t: {\cal E}_1 \to {\cal E}_0$ be 
two maps that make $1_{\cal E} \sim 0$, 
in the sense of the formula (\ref{homotopyformula}). 
Then we define two maps $F: {\cal F}\to C(f) $ and $G: C(f)\to {\cal F}$ by
\begin{displaymath}
\xymatrix{
{\cal F}_0  \ar@/_/[d]_{P_{\cal F}} \ar[r]^(.45){[1,0]^T} &
{\cal F}_0 \oplus {\cal E}_1 \ar@/_/[d]_{\hat{P}} \\
{\cal F}_1  \ar@/_/[u]_{Q_{\cal F}} \ar[r]^(.45){[1,0]^T}&
{\cal F}_1 \oplus {\cal E}_0 \ar@/_/[u]_{\hat{Q}}
}
\hspace{3cm}
\xymatrix@C=30pt{
{\cal F}_0 \oplus {\cal E}_1\ar@/_/[d]_{\hat{P}} \ar[r]^(.55){[1, f_0 t]} &
{\cal F}_0  \ar@/_/[d]_{P_{\cal F}}  \\
{\cal F}_1 \oplus {\cal E}_0 \ar@/_/[u]_{\hat{Q}} \ar[r]^(.55){[1,f_1 s]} &
{\cal F}_1  \ar@/_/[u]_{Q_{\cal F}}
}
\end{displaymath}
Then $GF = 1_{\cal F}$ exactly, and $FG \sim 1_{C(f)}$ using the maps
$$S \: = \: \left[ \begin{array}{cc}
0 & 0  \\
0 & -t 
\end{array} \right] : \;\; {\cal F}_0 \oplus {\cal E}_1 \:
\longrightarrow \: {\cal F}_1 \oplus {\cal E}_0  $$
$$T \: = \: \left[ \begin{array}{cc}
0 & 0  \\
0 & -s
\end{array} \right] : \;\; {\cal F}_1 \oplus {\cal E}_0 
\: \longrightarrow \: {\cal F}_0 \oplus {\cal E}_1  $$

Here is a a consequence of this result: 
suppose we have a short exact sequence of matrix factorizations
\begin{displaymath}
0 \: \longrightarrow \:
{\cal E} \: \stackrel{f}{\longrightarrow} \: {\cal F} \: 
\stackrel{g}{\longrightarrow} \: {\cal G} \: \longrightarrow \: 0
\end{displaymath}
and ${\cal E}$ is contractible.  Then, ${\cal G}$ is quasi-isomorphic
to ${\cal F}$.

To see this, recall the map $g$ above defines a map
$\tilde{g}:C(f) \to {\cal G} $
and $C(\tilde{g})$ is the iterated cone over the sequence. 
Then $C(\tilde{g})$ is homotopy-equivalent to 
(in fact, identical to) the iterated
cone over an exact sequence, hence $\tilde{g}$ is a quasi-isomorphism,
and ${\cal G}$ is quasi-isomorphic to $C(f)$ (by the definition of 
\textquoteleft quasi-isomorphism'), 
but $C(f)$ is homotopy-equivalent, and hence quasi-isomorphic, to ${\cal F}$.

\subsection{Duals, local Homs, and tensor products}
\label{sect:app:duals-homs}

Given $({\cal E},d_{\cal E}) = ({\cal E}_{0,1}, P_{\cal E}, Q_{\cal E})$ 
a matrix factorization
of a Landau-Ginzburg model with superpotential $W$, and 
$({\cal F},d_{\cal F}) = ({\cal F}_{0,1}, P_{\cal F}, Q_{\cal F})$
a matrix factorization of a Landau-Ginzburg model with superpotential
$W'$,
we shall define a dual $\vee$, local Hom,
and tensor products.

First, for ${\cal E}$ a matrix factorization
\begin{displaymath}
\xymatrix{
{\cal E}_0 \ar@/_/[d]_{P} \\
{\cal E}_1 \ar@/_/[u]_{Q}
}
\end{displaymath}
(of superpotential $W$),
the dual matrix factorization ${\cal E}^{\vee}$ is defined to be
\begin{displaymath}
\xymatrix{
{\cal E}_0^* \ar@/_/[d]_{-Q^*} \\
{\cal E}_1^* \ar@/_/[u]_{P^*}
}
\end{displaymath}
where
\begin{eqnarray*}
Q^*: \: {\cal H}om_{{\cal O}_X}({\cal E}_0, {\cal O}_X) & \longrightarrow &
{\cal H}om_{{\cal O}_X}({\cal E}_1, {\cal O}_X) \mbox{ as } \circ Q \\
P^*: \: {\cal H}om_{{\cal O}_X}({\cal E}_1, {\cal O}_X) & \longrightarrow &
{\cal H}om_{{\cal O}_X}({\cal E}_0, {\cal O}_X) \mbox{ as } \circ P
\end{eqnarray*}
It is easily checked that this is a matrix factorization of the
superpotential $-W$.

Next, we define ${\cal H}om({\cal E},{\cal F})$ to be the matrix factorization
\begin{displaymath}\label{2periodic}
\xymatrix{
 {\cal H}om_0({\cal E}, {\cal F}) \ar@/_/[d]_{\hat{P}} & = & {\cal H}om_{{\cal O}_X}({\cal E}_0, {\cal F}_0) \oplus
{\cal H}om_{{\cal O}_X}({\cal E}_1, {\cal F}_1) \ar@/_/[d]_{\hat{P}} \\
 {\cal H}om_1({\cal E}, {\cal F})\ar@/_/[u]_{\hat{Q}} &=& 
{\cal H}om_{{\cal O}_X}({\cal E}_0, {\cal F}_1) \oplus
{\cal H}om_{{\cal O}_X}({\cal E}_1, {\cal F}_0) \ar@/_/[u]_{\hat{Q}}
}
\end{displaymath}
where
\begin{displaymath}
\hat{P} \: = \: \left[ \begin{array}{cc}
P_{\cal F *} & -P_{\cal E}^* \\
-Q_{\cal E}^* & Q_{\cal F *}
\end{array} \right], \: \: \:
\hat{Q} \: = \: \left[ \begin{array}{cc}
Q_{\cal F*} & P_{\cal E}^* \\
Q_{\cal E}^* & P_{\cal F*}
\end{array} \right]
\end{displaymath}
This is a matrix factorization of the
Landau-Ginzburg model with superpotential $W'-W$.

Note that if $W'=0$ and we take ${\cal F}$ to be the matrix factorization
\begin{displaymath}
\xymatrix{
{\cal O}_X \ar@/_/[d] \\
0 \ar@/_/[u]
}
\end{displaymath}
then ${\cal H}om({\cal E},{\cal F})$ is the same matrix factorization of $-W$
as 
${\cal E}^{\vee}$.

Finally, let us define tensor products.
Given matrix factorizations ${\cal E}$, ${\cal F}$ as above,
define ${\cal E} \otimes {\cal F}$ to be the matrix factorization
\begin{displaymath}
\xymatrix{
    ({\cal E}_0 \otimes {\cal F}_0) \oplus ({\cal E}_1 \otimes {\cal F}_1)
\ar@/_/[d]_{\hat{P}} \\
({\cal E}_0 \otimes {\cal F}_1) \oplus ({\cal E}_1 \otimes {\cal F}_0)
 \ar@/_/[u]_{\hat{Q}}
}
\end{displaymath}
where
\begin{displaymath}
\hat{P} \: = \: \left[ \begin{array}{cc}
1 \otimes P_{\cal F} & -Q_{\cal E} \otimes 1 \\
 P_{\cal E} \otimes 1 & 1 \otimes Q_{\cal F}
\end{array} \right], \: \: \:
\hat{Q} \: = \: \left[ \begin{array}{cc}
1 \otimes Q_{\cal F} & Q_{\cal E} \otimes 1 \\
- P_{\cal E} \otimes 1 & 1 \otimes P_{\cal F}
\end{array} \right]
\end{displaymath}
This defines a matrix factorization the superpotential $W'+W$. More compactly, we can express the tensor product of $({\cal E}, d_{\cal E})$
and $({\cal F}, d_{\cal F})$ to be ${\cal E} \otimes {\cal F}$
with differential
\begin{displaymath}
d_{{\cal E}\otimes {\cal F}} \: = \: d_{\cal E}\otimes 1 \: + \:
1 \otimes d_{\cal F}
\end{displaymath}
with signs determined by the Koszul convention.

It is straightforward to show that ${\cal E} \otimes {\cal F}$
is isomorphic to ${\cal F} \otimes {\cal E}$, and also easy to check that 
${\cal E}^{\vee} \otimes {\cal F}$ is the same as 
${\cal H}om({\cal E}, {\cal F})$.

\subsection{Ext groups}
\label{app:Extgroups}

Next we need to discuss the space of massless open strings between two matrix 
factorizations, {\it i.e.} the morphisms in the category of 
topological B-branes. 

In the case that $W\equiv 0$, and so B-branes are described by ordinary complexes of sheaves, it is well known that the space of massless open strings between two branes is given by the Ext groups between the two complexes. When we have a non-zero $W$, and our branes are matrix factorizations, we can use a similar construction to define the space of massless open strings. We'll denote this space by
$${\rm Ext}^*_{MF}( {\cal E}, {\cal F}) $$
for ${\cal E}, {\cal F}$ a pair of matrix factorizations. Let's describe its construction.

To start with, let's assume we're working on an affine space $X$, and fix a superpotential $W$. If ${\cal E}$ and ${\cal F}$ are matrix factorizations of $W$, then (as described above) we can form ${\cal H}om({\cal E},{\cal F})$, which is a matrix factorization of $W-W=0$. If we take global sections of this, we get a $\mathbb{Z}_2$-graded chain complex of vector spaces

\begin{displaymath}
\xymatrix{
\Gamma_X {\cal H}om_0({\cal E}, {\cal F}) \ar@<.5ex>[r]^{\hat{P}}  &
 \Gamma_X{\cal H}om_1({\cal E}, {\cal F})\ar@<.5ex>[l]^{\hat{Q}}
}
\end{displaymath}

When we have a $\mathbb{C}^\times_R$ R-symmetry, we get something slightly better. The spaces of global sections split up into eigenspaces for the symmetry, so we actually have an honest ($\mathbb{Z}$-graded) chain-complex
$$ \dotsb \rightarrow (\Gamma_X {\cal H}om({\cal E}, {\cal F}))_0 \rightarrow (\Gamma_X {\cal H}om({\cal E}, {\cal F}))_1 \rightarrow (\Gamma_X {\cal H}om({\cal E}, {\cal F}))_2 \rightarrow \dotsb $$ 
In either case, the zeroeth homology of this chain-complex describes maps between ${\cal E}$ and ${\cal F}$, modulo homotopies.

As long as ${\cal E}$ and ${\cal F}$ are traditional matrix factorizations built out of vector bundles (and we continue to assume that $X$ is affine) then we are basically done: the homology of the above chain complex defines ${\rm Ext}^*_{MF}( {\cal E}, {\cal F})$. However, for sheafy matrix factorizations the recipe is more complicated, let's break it down into steps:
\begin{enumerate}
\item  Find a traditional (vector bundle) matrix factorization $\hat{\mathcal{E}}$ that is quasi-isomorphic to ${\cal E}$.
\item  Form ${\cal H}om(\hat{{\cal E}}, {\cal F})$, a matrix factorization of 0.
\item Take global sections to get $\Gamma_X {\cal H}om(\hat{{\cal E}}, {\cal F})$, a chain-complex of vector spaces.
\item Take homology of this chain-complex to get ${\rm Ext}^*_{MF}( {\cal E}, {\cal F})$.\footnote{For some purposes it is appropriate to stop at step 3, 
and declare that the set of morphisms between two matrix factorizations is 
actually a chain-complex (rather than its homology). This makes the category 
of matrix factorizations a dg-category.} 
\end{enumerate}

This recipe is very closely analogous to the procedure for computing Ext groups between sheaves in the usual derived category, in particular we should think of the first step as finding a `projective resolution' of ${\cal E}$. Given this analogy, it is reasonable to use the notation
$$ {\rm R}{\cal H}om_{MF}({\cal E}, {\cal F}) := {\cal H}om(\hat{{\cal E}}, {\cal F})$$
for the matrix factorization obtained in step 2. Also, because this is a matrix factorization of 0, it makes sense to take its homology, which we denote by
$${\cal E}xt_{MF}( {\cal E}, {\cal F}) $$
This object consists of a pair of sheaves, or, in the presence of a $\mathbb{C}^\times_R$ R-symmetry, a single sheaf with a $\mathbb{C}^\times_R$ action. It is analogous to the local ${\cal E}xt$ sheaves that one computes in the ordinary derived category.

We will work out specific examples of this recipe in 
sections~\ref{sect:RG:functor},
\ref{sect:onequadric}, and appendix~\ref{app:pointlike}.

Now let's drop the assumption that $X$ is affine. In fact, most of the calculations that we do in this paper take place in local (affine) models, so the technical details of handling non-affine spaces are not so important for us. Nevertheless we will make a few claims about non-affine examples, so let's say a few words about them.

We see immediately that in the non-affine case the above recipe is not sufficient - even if we set $W\equiv 0$, and let ${\cal E}$ and ${\cal F}$ be single vector bundles, then the above recipe outputs $\Gamma_X(\mathcal{H}om(\mathcal{E}, \mathcal{F}))$, whereas the correct space of massless open strings is actually 
$${\rm Ext}_X(\mathcal{E}, \mathcal{F}) = {\rm H}^*_X( \mathcal{H}om(\mathcal{E}, \mathcal{F}) )$$
The implication of this is that we have to do something more sophisticated in step 3 of the recipe,
 instead of taking just global sections of ${\rm R}{\cal H}om_{MF}({\cal E}, {\cal F})$ we have to take its \emph{derived} global sections
$$ {\rm R}\Gamma_X {\rm R}{\cal H}om_{MF}({\cal E}, {\cal F})$$
This means we have to write down a chain-complex that computes the cohomology of the sheaf underlying ${\rm R}{\cal H}om_{MF}({\cal E}, {\cal F})$, which we can do by using \v{C}ech resolutions, or (if ${\cal F}$ is a vector bundle) Dolbeaut resolutions. Then we perturb the differential on this chain-complex by adding in the differential that acts on ${\rm R}{\cal H}om_{MF}({\cal E}, {\cal F})$. Then ${\rm Ext}^*_{MF}( {\cal E}, {\cal F})$ is the homology of this final complex.

This is rather complicated, and fortunately in practice we can usually do something simpler. Instead of writing down ${\rm R}\Gamma_X$ of the sheaf ${\rm R}{\cal H}om_{MF}({\cal E}, {\cal F})$, we just write down its cohomology groups
$$ {\rm H}^*_X ({\rm R}{\cal H}om_{MF}({\cal E}, {\cal F}))$$
and add in the differential. 
If we are lucky ({\it i.e.} if the relevant spectral sequence collapses) then 
this chain complex correctly computes ${\rm Ext}^*_{MF}( {\cal E}, {\cal F})$.

In many cases we can simplify even further, by using the `local-to-global' 
spectral sequence. If we take the homology of 
${\rm R}{\cal H}om_{MF}({\cal E}, {\cal F})$ before we take global sections, 
then we have a spectral sequence beginning with
$${\rm H}^*_X ({\cal E}xt_{MF}({\cal E}, {\cal F}))$$
and converging to ${\rm Ext}^*_{MF}( {\cal E}, {\cal F})$.

Let's apply this technology to a useful class of examples. Suppose we have a Landau-Ginzburg model over $X$, and we have a submanifold $Y$ lying within the locus $\{W = 0 \}$. Then the skyscraper sheaf ${\cal O}_Y$ defines a sheafy matrix factorization, and we can attempt to compute 
${\rm Ext}^*_{MF}( {\cal O}_Y, {\cal O}_Y)$. The first step is to find a traditional matrix factorization that is quasi-isomorphic to ${\cal O}_Y$.

Let's assume $Y$ is a complete intersection, cut out by sections $s_1,\dotsc,s_r$ of line-bundles $L_1,\dotsc,L_r$. Then the sheaf ${\cal O}_Y$ is resolved by a Koszul complex
of the form
\begin{displaymath}
0 \: \longrightarrow \:
\textstyle\bigwedge^r \mathcal{L}^* \: \longrightarrow \: \cdots \:
\longrightarrow \: \bigwedge^k {\cal  L}^* \: 
\longrightarrow \: \cdots \: \longrightarrow \: 
{\cal L}^* \: \longrightarrow \: {\cal O} \: 
\longrightarrow \: 0
\end{displaymath}
where ${\cal L} = \oplus L_i$, and the differentials are given by contracting with $(s_1, \cdots, s_r)$,
as usual for a Koszul resolution.
Since $W|_Y = 0$, $W$ is in the ideal generated by the $s_i$, so we can
find sections $f_1,\dotsc,f_r$ of the dual line-bundles  $L^*_1,\dotsc,L^*_r$ such that
$$W = \sum_i f_i s_i$$
 If we add `backwards' arrows to the Koszul complex that wedge with $(f_1, \cdots, f_r)$, and
fold into a two-term complex, then we get a matrix factorization of $W$ 
(we will see an explicit example of this in section~\ref{sect:quasi-rgflow}).

Now we claim that the matrix factorization just constructed is
quasi-isomorphic to ${\cal O}_Y$.  To show this, we begin with the case $r=1$.
We have an exact sequence of matrix factorizations
\begin{displaymath}
\xymatrix{
{\cal L}^* \ar[r]^s \ar@/_/[d]_W & 
{\cal O} \ar[r] \ar@/_/[d]_f & {\cal O}_Y \\
{\cal L}^* \ar[r]^1 \ar@/_/[u]_1 
& 
{\cal L}^* \ar@/_/[u]_s
}
\end{displaymath}
The leftmost matrix factorization is contractible, so the result follows.
When $r>1$, take a copy of the left-hand square for each triple $s_i, f_i, L_i$,
and tensor them together.  This produces an exact sequence of matrix 
factorizations resolving ${\cal O}_Y$, and all past the first are
contractible because they contain at least one contractible factor. And the first one is the `perturbed Koszul' matrix factorization described above.

Now we can compute ${\rm R}{\cal H}om_{MF}({\cal O}_Y, {\cal O}_Y)$. It has the form
\begin{displaymath}
0 \: \longleftarrow \:{\cal O}_Y \: \longleftarrow \: 
{\cal L}_Y  \: \longleftarrow \: \cdots \:
\longleftarrow \: \textstyle\bigwedge^k {\cal  L}_Y \: 
\longleftarrow\: \cdots \: \longleftarrow \: \bigwedge^r \mathcal{L}_Y \: 
\longleftarrow \: 0
\end{displaymath}
where the maps are given by contracting with $(f_1,\dotsc,f_r)$ (the $s_i$ have gone to zero, since it's supported on $Y$). There's a more invariant way to say this - the bundle ${\cal L}_Y$ on $Y$ is actually the normal bundle $N_{Y/X}$, and the differential is actually `contract with $dW$'. This makes sense because there is a well-defined map
\begin{displaymath}
dW: \: N_{Y/X} \: \longrightarrow \: {\cal O}_Y
\end{displaymath}
Consequently, ${\rm Ext}^*_{MF}( {\cal O}_Y, {\cal O}_Y)$ is computed from a 
spectral sequence beginning with $${\rm H}^*_Y(\textstyle\bigwedge N_{Y/X})$$
and the differential given by contracting with $dW$. 

Writing the result in this invariant form, it is clear that assuming that $Y$ is a complete intersection is actually not necessary, this result holds for any submanifold $Y$ in $\{W=0\}$. Also, notice that if we view $\mathcal{O}_Y$ as an object in the ordinary derived category, we have
 $${\rm Ext}^*_{X}( {\cal O}_Y, {\cal O}_Y) ={\rm H}^*_Y(\textstyle\bigwedge N_{Y/X}) $$
so the result we obtain in the category of matrix factorizations is a deformation of the result in the ordinary derived category.

This calculation generalizes the BRST cohomology results
discussed in {\it e.g.} \cite{ks}, in that adding a superpotential
deforms the BRST operator in the B-twisted theory by adding contractions
with $dW$.  This was discussed in \cite{alg22}, and used there to observe
that the closed string spectrum is computed by hypercohomology of a complex
generated by contractions with $dW$. The calculation here is the analogous result for open strings.

\section{Point-like behavior of ${\cal O}_U$}
\label{app:pointlike}

In this appendix we perform an explict calculation in a very simple Landau-Ginzburg model. This calculation is a key point in our analyses of the more complex models considered in section~\ref{sect:smooth}.

Let $V$ be a vector space, and let $W$ be a (possibly degenerate) quadratic form on $V$. We're going to work going to work in the Landau-Ginzburg model defined on $V$, with superpotential $W$. We're going to assume that the rank of $W$ is even  (the odd-rank case can be treated by the same method, but the answer turns out to be slightly different). In this case we can choose 
coordinates $x_1,\cdots,x_k,y_1,\cdots,y_k,z_1,\cdots,z_m$ such that 
\begin{displaymath}
W \: = \: x_1 y_1 \: + \: \cdots \: + \:
x_n y_n
\end{displaymath}
We let $Z$ denote the kernel of $W$, it is the subspace spanned by the 
$z$-coordinates.

We let $U$ be the $k$-dimensional subspace
\begin{displaymath}
U \: = \: \{ x_1 \: = \: \cdots \: = \: x_k \: = z_1 \: = \: \cdots \: = \: z_m \: =\: 0 \}
\end{displaymath}
Then $U$ is isotropic, has trivial intersection with $Z$, and has maximal rank among subspaces obeying these two conditions. Any other subspace with this property is equivalent to $U$ under a change-of-coordinates. In many of our applications $W$ is actually non-degenerate, in which case $Z=0$ and $U$ is just a maximal isotropic subspace.

Since $W$ vanishes along the subspace $U$, the sheaf ${\cal O}_U$ defines a sheafy matrix factorization
\begin{displaymath}
\xymatrix{
{\cal O}_U \ar@/_/[d] \\
0 \ar@/_/[u]
}
\end{displaymath}
The important result for us is that this sheafy matrix factorization is in fact pointlike, {\it i.e.} it may be viewed as a D0-brane. 
We are going to describe the arguments for this in detail in this appendix.

Let's compute ${\rm R}{\cal H}om_{MF}({\cal O}_U, {\cal O}_U)$. 
In fact we have already discussed this calculation in greater generality in 
appendix~\ref{app:Extgroups}, but we'll give more details here for this 
specific example. The first step is to find a traditional matrix factorization 
quasi-isomorphic to ${\cal O}_U$, which we do using Koszul resolutions.

Let $U^\perp\subset V$ be the subspace spanned by the $x$ and $z$ coordinates. For notational convience, we'll relabel the coordinates on $U^\perp$ as
$$ \tilde{x}_1 = x_1,\;\;\cdots,\;\; \tilde{x}_k=x_k,\;\; 
\tilde{x}_{k+1}=z_1,\;\; \cdots,\;\; \tilde{x}_{k+m}=z_m $$
and let  $\{ e_1,\dotsc,e_{k+m} \}$ be the corresponding basis vectors for 
$U^\perp$. Then one can write the Koszul resolution of ${\cal O}_U$ on $V$ as
\begin{displaymath}
\cdots \: \stackrel{\partial}{\longrightarrow} \: 
{\cal O}_V^{k+m \choose 3}
\: \stackrel{\partial}{\longrightarrow} \:
{\cal O}_V^{k+m \choose 2} 
\: \stackrel{\partial}{\longrightarrow} \:
{\cal O}_V^{k+m \choose 1}
\: \stackrel{\partial}{\longrightarrow} \:
{\cal O}_V \: \longrightarrow \: {\cal O}_U \: \longrightarrow \: 0
\end{displaymath}
where each term
\begin{displaymath}
{\cal O}_V^{k+m \choose p}
\: = \: {\cal O}_V \otimes_{\mathbb C}
{\mathbb C}\{ e_{i_1} \wedge \cdots \wedge e_{i_p} \}
\end{displaymath}
and the maps act as
\begin{displaymath}
\partial \left( e_{i_1} \wedge \cdots \wedge e_{i_p} \right) \: = \:
\sum_{r=1}^p (-)^{r-1} \tilde{x}_{i_r} e_{i_1} \wedge \cdots \wedge
\widehat{ e_{i_r} } \wedge \cdots \wedge e_{i_p}
\end{displaymath}
(for $p=1$, this says that $\partial e_i = \tilde{x}_i$).  It is a standard result
that $\partial^2 = 0$.  For example, if $k=2$ and $m=0$, this becomes the sequence
\begin{displaymath}
0 \: \longrightarrow \: {\cal O}_V \:
\stackrel{ [-x_2, x_1]^T 
}{\longrightarrow} \: {\cal O}_V^2 \: 
\stackrel{ [x_1, x_2] }{\longrightarrow} \: {\cal O}_V 
\: \longrightarrow \: {\cal O}_U \: \longrightarrow \: 0
\end{displaymath}

Next, we define maps $\delta$ going in the opposite direction:
\begin{displaymath}
\cdots \: \stackrel{\delta}{\longleftarrow} \:
{\cal O}_V^{k+m \choose 3}
\: \stackrel{\delta}{\longleftarrow} \:
{\cal O}_V^{k+m \choose 2}
\: \stackrel{\delta}{\longleftarrow} \:
{\cal O}_V^{k+m \choose 1}
\: \stackrel{\delta}{\longleftarrow} \:
{\cal O}_V
\end{displaymath}
by
\begin{displaymath}
\delta\left( e_{i_1} \wedge \cdots \wedge e_{i_p} \right) \: = \:
\sum_{j=1}^k y_j e_j \wedge e_{i_1} \wedge \cdots \wedge e_{i_p}
\end{displaymath}

To define the matrix factorization that is quasi-isomorphic to ${\cal O}_U$,  we use both the maps $\partial$ and $\delta$. Schematically, it's given by
\begin{displaymath}
\xymatrix{
 {\cal O}_V^{k+m \choose {\rm even}}\ar@/_/[d]_{\partial + \delta} \\
{\cal O}_V^{k+m \choose {\rm odd}} \ar@/_/[u]_{\partial + \delta}
}
\end{displaymath}

To make this more clear, let us consider the special case that $n=2$, $m=0$.
In this case, we have a matrix factorization
\begin{displaymath}
\xymatrix{ 
{\cal O}_V^2 \: = \: {\cal O}_V \otimes_{\mathbb C} 
{\mathbb C}\{ 1, e_1 \wedge e_2 \} \ar@/_/[d] \\
{\cal O}_V^2 \: = \: {\cal O}_V \otimes_{\mathbb C}
{\mathbb C}\{ e_1, e_2 \} \ar@/_/[u]
}
\end{displaymath}
Since
\begin{eqnarray*}
\partial(e_1, e_2) & = & (x_1, x_2)(1) \\
\delta(e_1, e_2) & = & (-y_2, y_1) e_1 \wedge e_2
\end{eqnarray*}
we find that
\begin{displaymath}
\uparrow \: = \: \left[ \begin{array}{cc}
x_1 & x_2 \\
-y_2 & y_1 
\end{array} \right].
\end{displaymath}
Similarly, since
\begin{eqnarray*}
\delta(1) & = & y_1 e_1 \: + \: y_2 e_2 \\
\partial(e_1 \wedge e_2) & = & -x_2 e_1 \: + \: x_1 e_2
\end{eqnarray*}
we find that
\begin{displaymath}
\downarrow \: = \: \left[ \begin{array}{cc}
y_1 & -x_2 \\
y_2 & x_1 \end{array} \right]
\end{displaymath}
It is easy to check that $\uparrow \circ \downarrow = W \, {\rm Id} = 
\downarrow \circ \uparrow$.

More concisely, if we define
\begin{displaymath}
{\cal E} \: = \: {\cal O}_V \otimes \textstyle\bigwedge^*(U^\perp)
\end{displaymath}
then the matrix factorization above can be described by the
single differential $d$, given by
\begin{displaymath}
d(f \otimes \omega) \: = \: \sum_{i=1}^{k+m} 
\tilde{x}_i f \otimes {e_i}{\lrcorner} \omega  \: +\: \sum_{i=1}^k y_i f \otimes e_i \wedge \omega
\end{displaymath}
where $f \in \Gamma({\cal O}_V)$, $\omega \in \textstyle\bigwedge^*(U^\perp)$.
It is then straightforward to check that $d^2 = W \, {\rm Id}$.

Now we are ready to compute Ext groups and verify pointlike behavior.  Taking ${\cal H}om$ from
the matrix factorization above to ${\cal O}_U$, we get the complex
\begin{displaymath}
\xymatrix{
 {\cal O}_U^{k+m \choose {\rm even}}\ar@/_/[d]_{ \delta} \\
{\cal O}_U^{k+m \choose {\rm odd}} \ar@/_/[u]_{\delta}
}
\end{displaymath}
(since $\partial =0$ along $U$), or, writing it in an `unrolled' way,
\begin{displaymath}
{\cal O}_U
\: \stackrel{\delta}{\longleftarrow} \:
{\cal O}_U^{k+m \choose 1}
\: \stackrel{\delta}{\longleftarrow} \:
\cdots 
\: \stackrel{\delta}{\longleftarrow} \:
{\cal O}_U^{k+m \choose 2}
\: \stackrel{\delta}{\longleftarrow} \:
{\cal O}_U^{k+m \choose 1}
\: \stackrel{\delta}{\longleftarrow} \:
{\cal O}_U
\end{displaymath}
In the case that $W$ is non-degenerate, so $m=0$, this complex is precisely the Koszul resolution of the origin in $U$. So in this case we can conclude that 
$${\cal E}xt_{MF}({\cal O}_U, {\cal O}_U) = {\cal O}_0 $$
(the sky-scraper sheaf at the origin), and  hence
$${\rm Ext}^0_{MF}({\cal O}_U, {\cal O}_U) = \mathbb{C},\;\;\;\; {\rm Ext}^{\neq 0}_{MF}({\cal O}_U, {\cal O}_U) = 0 $$
Thus in the non-degenerate case ${\cal O}_U$ is homologically point-like, where the `point' sits in a zero-dimensional space.

Now let's allow $m>0$. With a little more work, one can see that the homology of the above complex gives
$${\cal E}xt_{MF}({\cal O}_U, {\cal O}_U) = {\cal O}_0 
\:\oplus\:
{\cal O}_0^{m \choose 1}
\: \oplus \:
{\cal O}_0^{m \choose 2}
\: \oplus \:
\cdots
\: \oplus \:
{\cal O}_0^{m \choose 1}
\: \oplus \:
{\cal O}_0 
$$
and  hence
$${\rm Ext}^p_{MF}({\cal O}_U, {\cal O}_U) = \mathbb{C}^{m \choose p} $$
So in general ${\cal O}_U$ is homologically point-like, where the `point' sits in an $m$-dimensional space. This $m$-dimensional space is of course the kernel $Z$, indeed if we take a little more care with all our vector spaces we can see that in fact
$${\rm Ext}^p_{MF}({\cal O}_U, {\cal O}_U) = \textstyle\bigwedge^p Z $$
These are canonically equal to the self-Ext groups of the sky-scraper sheaf supported at the origin in $Z$.

Now let's discuss the criterion of set-theoretic support. Given the calculation we've just performed, it will come as no surprise if we claim that the set-theoretic support of ${\cal O}_U$ is the origin in $V$. 
In fact our calculation implies this precisely, since if we delete the origin 
from $V$ then ${\cal E}xt_{MF}({\cal O}_U, {\cal O}_U)$ goes to zero 
(since it was supported at the origin), and consequently so does 
${\rm Ext}^0_{MF}({\cal O}_U, {\cal O}_U)$. But this implies in particular 
that the identity map on ${\cal O}_U$ must be exact, {\it i.e.} it must be 
homotopic to the zero map. 
So away from the origin, ${\cal O}_U$ is contractible. 

As a final remark, let us observe that we are free to put any (compatible) R-symmetry on this model, or to replace $V$ by the orbifold $[V/\mathbb{Z}_2]$, and the calculation given here will go through unchanged. In the non-degenerate case it will not even affect the answer, since the origin in $V$ will be fixed by any R-symmetry or orbifold action.

\end{document}